\documentclass[preprint,12pt,authoryear]{elsarticle}

\usepackage{amsmath}
\usepackage{amssymb}
\usepackage{xcolor}
\usepackage{hyperref}
\hypersetup{colorlinks=true}

\newcommand{\EE}{\mathbb{E}}
\newcommand{\PP}{\mathbb{P}}
\newcommand{\II}{\mathbb{I}}
\newcommand{\dif}{\mathrm{d}}
\newcommand{\mask}{\texttt{[mask]}}
\newcommand{\hD}{\hat{\Delta}}
\newcommand{\tin}{t_{\mathrm{in}}}
\newcommand{\minit}{m_{\mathrm{in}}}
\newcommand{\mout}{m_{\mathrm{out}}}
\newcommand{\Pmem}{P_{\mathrm{mem}}}
\newcommand{\leff}{\lambda_{\mathrm{eff}}}
\newcommand{\iid}{\overset{\mathrm{iid}}{\sim}}
\DeclareMathOperator{\Var}{Var}
\DeclareMathOperator{\Cov}{Cov}
\DeclareMathOperator{\softmax}{softmax}
\DeclareMathOperator{\prox}{prox}
\DeclareMathOperator*{\extr}{extr}
\DeclareMathOperator*{\argmin}{arg\,min}
\DeclareMathOperator{\sign}{sign}
\DeclareMathOperator{\sech}{sech}
\begin{document}

\begin{frontmatter}

\title{A Minimal Masked Diffusion Language Model as an Associative Memory}

\author[bocconi]{Filippo Elgorni\corref{cor1}}
\ead{filippo.elgorni@phd.unibocconi.it}
\cortext[cor1]{Corresponding author.}

\author[bocconi,bidsa]{Carlo Lucibello}
\ead{carlo.lucibello@unibocconi.it}

\affiliation[bocconi]{organization={Department of Computing Sciences, Bocconi University},
            city={Milan},
            country={Italy}}

\affiliation[bidsa]{organization={Bocconi Institute for Data Science and Analytics, Bocconi University},
            city={Milan},
            country={Italy}}

\begin{abstract}
Masked diffusion language models (MDLMs) have recently emerged as a flexible class of generative models for discrete data, extending autoregressive approaches by enabling any-order generation and the parallel unmasking of multiple tokens.
As for continuous diffusion models, a central question regards understanding under which conditions these models memorize their training data rather than generalize beyond it.

We first provide empirical evidence that memorization phenomena are indeed present in practical MDLMs trained on small subsets of real datasets.
Motivated by this observation, we introduce a minimal and analytically tractable model of a trained masked diffusion system, in which the denoiser is a single dense layer and the training data are uniformly distributed binary sequences.
In the high-dimensional proportional limit, we show that this model naturally behaves as an associative memory, acting as a retrieval system under partial observation.

Using the replica method, we characterize the trained weights as a function of the load $\alpha$. We derive the exact retrieval performance of one-step decoding and a low-dimensional dynamical description that is approximate but accurate for step-by-step decoding.
These theoretical predictions are validated by numerical experiments.
\end{abstract}


\end{frontmatter}

\section{Introduction}

Autoregressive (AR) Transformer-based neural networks constitute the backbone of current state-of-the-art large language models (LLMs) \citep{vaswani2017attention,brown2020language}.
Autoregressive generation comes with the inherent limitation that tokens must be generated sequentially, and $L$ forward passes of the model, each one taking in the full previously generated context, are needed to generate a sequence of length $L$, with a computational cost that scales accordingly.
This is partially mitigated by KV caching \citep{pope2023efficiently}.
Several alternatives to the AR paradigm have been proposed for speeding up discrete sequence generation,
with the two most prominent for language being masked diffusion models, operating fully in discrete space, and diffusion and flow models with continuous latents \citep{li2022diffusion}.

Masked Diffusion Language Models (MDLMs) \citep{austin2021structured,lou2023discrete,campbell2022continuous,sahoo2024simple,shi2024simplified} have already been successfully applied
to natural language and code generation \citep{nie2026large,inception2025mercury}, biological sequence modeling and protein design \citep{sahoo2024simple,wang2024diffusion}, discrete image generation \citep{shi2024simplified}, and sampling from energy-based distributions \citep{zhu2026mdns}.
In this sense, they are any-order autoregressive models \citep{ou2025your}, an interpretation that complements the noising--denoising picture of generative diffusion in which they were originally formulated.
Accelerated generation is possible because multiple tokens can be sampled in parallel, at the cost of discarding some correlations \citep{ben2026accelerated,lavenant2025error,secchi2026schedule,wainwright2026data}.

Most families of generative models, including autoregressive models and both continuous and discrete diffusion models, are prone to memorization of their training datasets. While this can be desirable in knowledge-retrieval tasks, it is detrimental when the goal is to generate novel samples, and it raises privacy and copyright concerns \citep{carlini2021extracting,carlini2023extracting,somepalli2023diffusion}.
One interpretation of this behaviour lies in the nature of the commonly used training objectives, which pull the models toward the empirical distribution of the training data when cued with partial observation of a training point \citep{biroli2024dynamical,achilli2025memorization,ventura2025manifolds}.
This behavior essentially resembles that of associative memories, a connection that has recently attracted attention \citep{ambrogioni2024search,d2026pseudo,pham2026language,pham2026emergence}.

The prototypical model of associative memory is the Hopfield Model (HM) \citep{hopfield1982neural}, a Gibbs measure whose energy has pairwise interactions with values prescribed by the Hebbian rule (although several training approaches also exist; \citealp{kanter1987associative,agliari2019dreaming,benedetti2022supervised,serricchio2025daydreaming}).
A pattern is said to be retrieved if a low-temperature dynamics initialized close to it remains in its basin of attraction.
While the dynamics of the HM is difficult to analyze exactly, a static computation using the replica method \citep{amit1985storing} reveals that $M$ patterns of length $L$ can be retrieved as long as the load $\alpha=M/L$ is smaller than a critical value $\alpha_c \approx 0.138$ as both $L$ and $M$ go to infinity.
In this paper we provide an analytical characterization of the memorization capacity of a simple MDLM in the same high-dimensional limit.

Our setting differs in some relevant aspects from that of the standard HM of associative memory \citep{hopfield1982neural}.
To start, the HM and MDLMs correspond to two different corruption models for the data.
A corrupted pattern in the former corresponds to an initial condition with some finite overlap $\minit$ with the target pattern, where the correct bits are in unknown positions and where the dynamics updates all the spins and has no fixed time horizon.
In contrast, in MDLMs a corrupted pattern corresponds to a partially observed pattern, where the observed bits are in known positions and the unobserved bits are to be inferred.
The dynamics unmasks the unobserved bits sequentially, and once a masked bit is assigned a value, it is no longer updated, therefore the retrieval process in MDLMs has a fixed time horizon.
Moreover, MDLMs have to be trained on the data before they can be used.

Our simple discrete score model is implemented as a single linear layer, where both the training and the retrieval dynamics turn out to be analytically tractable.
Linear score models \citep{biroli2023generative,merger2025generalization} and random-feature score models \citep{george2026denoising,bonnaire2026diffusion,urfin2026double} have been studied in continuous diffusion, where the linear score can only represent Gaussian distributions and thus multimodal distributions from the training set cannot be memorized.
Interestingly, this limitation is not present in the discrete case.

This paper is structured as follows.
\begin{itemize}
  \item In Section~\ref{sec:prel} we give a brief overview of MDLMs, and show that a realistic model trained on small subsets of CIFAR-10 reproduces its training images, both when cued with a heavily masked training image and when generating from scratch.
  \item In Section~\ref{sec:lsm} we introduce the linear score model (LSM), a single-layer masked diffusion model, and in Section~\ref{sec:replica} we characterize its weights trained on random binary patterns with the replica method, including the full distribution of the pattern stabilities.
  \item In Section~\ref{sec:retrieval} we compute exactly the retrieval performance of one-step decoding. For step-by-step decoding, we derive an exact Markovian closure for Hebbian couplings, both with a direct argument and with the Martin--Siggia--Rose formalism, and an approximate closure for the learned couplings, based on a Hebbian-like decomposition of the weights, which accurately matches simulations.
  We also show that the perfect retrieval of a pattern from a finite fraction of its tokens is impossible for $M$ linear in $L$.
  \item In Section~\ref{sec:pinned} we compare the LSM with a HM in which the observed spins are pinned to the pattern.
\end{itemize}

\section{Masked Diffusion Language Models}\label{sec:prel}

\subsection{Preliminaries}

The goal of MDLMs is to sample from a distribution $p_0$ defined over sequences $x \in \mathcal{V}^L$, where $\mathcal{V}$ is a vocabulary of size $V$.
The model is trained to predict individual tokens conditioned on arbitrary subsets of the other tokens: this is achieved by randomly masking some positions in the input sequence and training the model to predict the masked tokens from the unmasked ones.

\begin{figure}[htbp]
  \centering
  \includegraphics[width=0.7\linewidth]{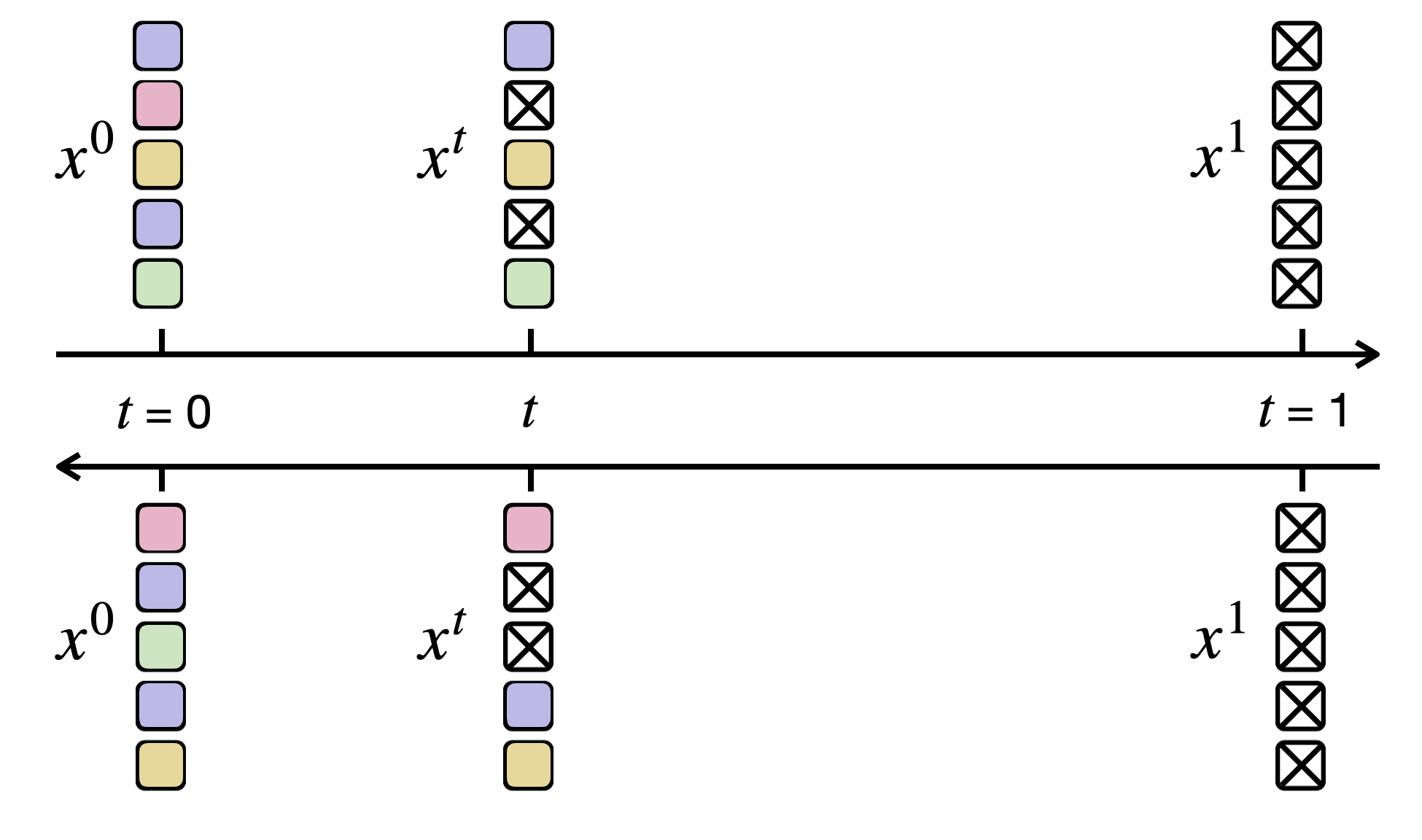}
  \caption{Masking (top) progressively replaces tokens with masks as time increases from $0$ to $1$.
    Sampling (bottom) starts from the fully masked sequence and progressively fills masked positions using the learned conditional distributions.
    Colors represent token values, and crossed boxes denote masks.}
  \label{fig:forwardbackward}
\end{figure}

Given a clean sequence $x^0 \sim p_0$, the forward masking process progressively corrupts it as time increases from $t=0$ to $t=1$ (Fig.~\ref{fig:forwardbackward}, top).
At time $t$, the corrupted sequence $x^t$ is obtained by independently replacing each token with a special mask token $\mask$ with probability $t$.

The masked diffusion model is a probabilistic model $p_\theta(x_i|x^t)$ that takes as input a masked sequence $x^t \in (\mathcal{V} \cup \{\mask\})^L$ and outputs probabilities over the tokens in $\mathcal{V}$ for each masked position $i$ where $x_i^t = \mask$.
Training minimizes the masked cross-entropy objective corresponding to the evidence lower bound \citep{sahoo2024simple,shi2024simplified}:
\begin{equation}
  \mathcal{L}(\theta) = -\EE_{t\sim U(0,1)} \frac{1}{t}\, \EE_{x^0,x^t} \sum_{i\,:\,x^t_i=\mask} \log p_\theta(x^0_i|x^t).
  \label{eq:loss}
\end{equation}

Once the model is trained, the backward sampling process proceeds in the opposite direction, from $t=1$ to $t=0$, using the learned conditional distributions to progressively unmask tokens (Fig.~\ref{fig:forwardbackward}, bottom).
This is implemented by discretizing time and selecting masked positions in random order starting from a fully masked sequence.
Notice that the sampling conditionals for each step can be obtained from $p_\theta(\cdot | x^t)$ with different schemes.

Because the model predicts single-token conditional distributions given arbitrary observed subsets, it supports autoregressive generation in any order.
Multiple positions can also be filled in parallel using the same partially masked sequence, speeding up generation at the cost of neglecting conditional correlations between the simultaneously generated tokens.

\subsection{Memorization Experiments}
We trained the masked diffusion model of \citet{schiff2025simple} on progressively smaller subsets of the CIFAR-10 dataset \citep{krizhevsky2009tiny} and observed phenomena compatible with memorization.
We worked with images instead of text to be able to work with a reduced vocabulary.
Each image is a sequence of $L = 32 \times 32 \times 3 = 3072$ tokens, one per color channel of each pixel, each taking one of 256 intensity values or the mask value.

\begin{figure}[htbp]
  \centering
  \includegraphics[width=\linewidth]{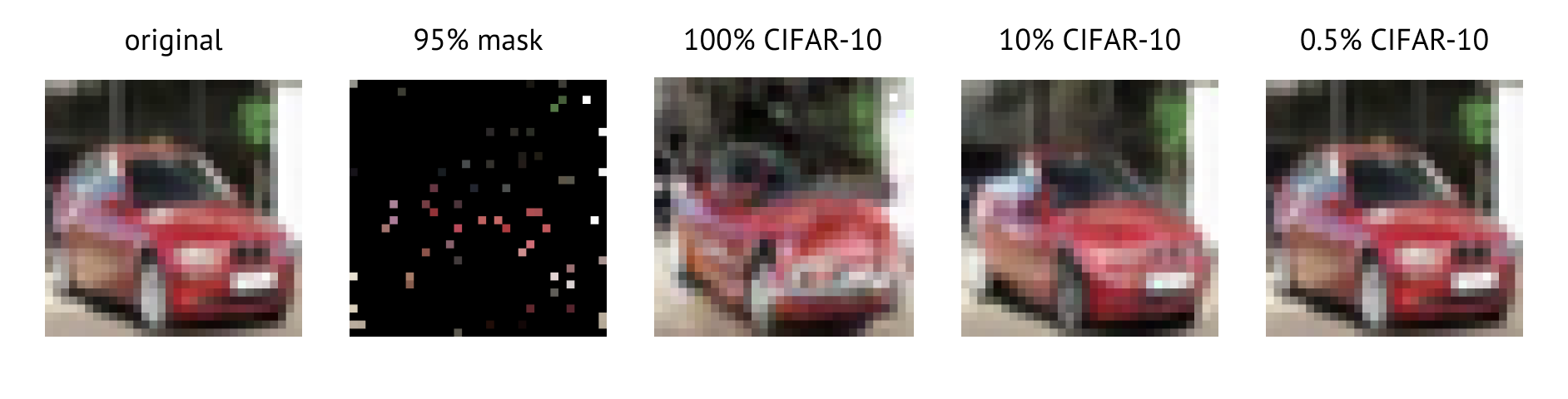}
  \caption{Reconstruction experiment on CIFAR-10. A training image (first panel) is masked at 95\% of its tokens (second panel) and then decoded by models trained on different fractions of the training set (remaining panels).
    In order to generate realistic images in the generalization regime, we used classifier-free guidance \citep{ho2022classifier,schiff2025simple} with scale $\gamma=2$.
    The number of decoding steps equals the number of tokens, $L = 3072$.
    Each model has been trained for 200000 steps.}
  \label{fig:uturn-cifar}
\end{figure}

We find that, for smaller datasets, the model is able to reconstruct the training images with high fidelity, even when starting from a version with as many as 95\% of its tokens masked, see Fig.~\ref{fig:uturn-cifar}.
Moreover, even when new images are generated from scratch, we find that the models trained on smaller datasets generate samples close (or in some cases nearly identical) to those of the original dataset in the $L_2$ sense.
We quantify memorization as proposed in \citet{yoon2023memorize}: a generated image $x$ is considered memorized if, denoting by $\xi^{\mu_1}$ and $\xi^{\mu_2}$ its nearest and second nearest neighbors in the training set, one has
\begin{equation}
  \frac{\|x-\xi^{\mu_1}\|_2}{\|x-\xi^{\mu_2}\|_2} < \kappa.
  \label{eq:ratio}
\end{equation}

\begin{figure}[htbp]
  \centering
  \includegraphics[width=0.49\linewidth]{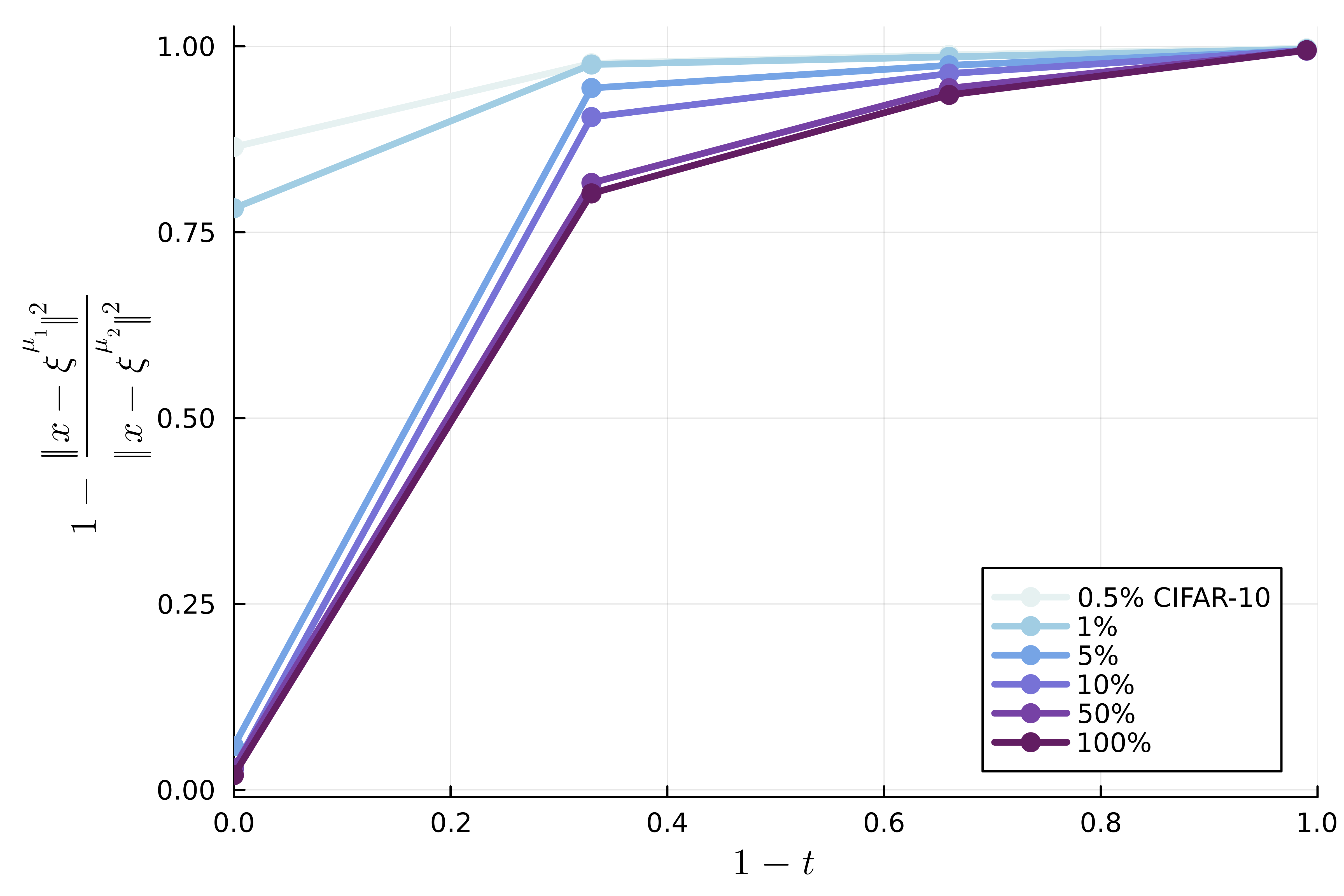}\hfill
  \includegraphics[width=0.49\linewidth]{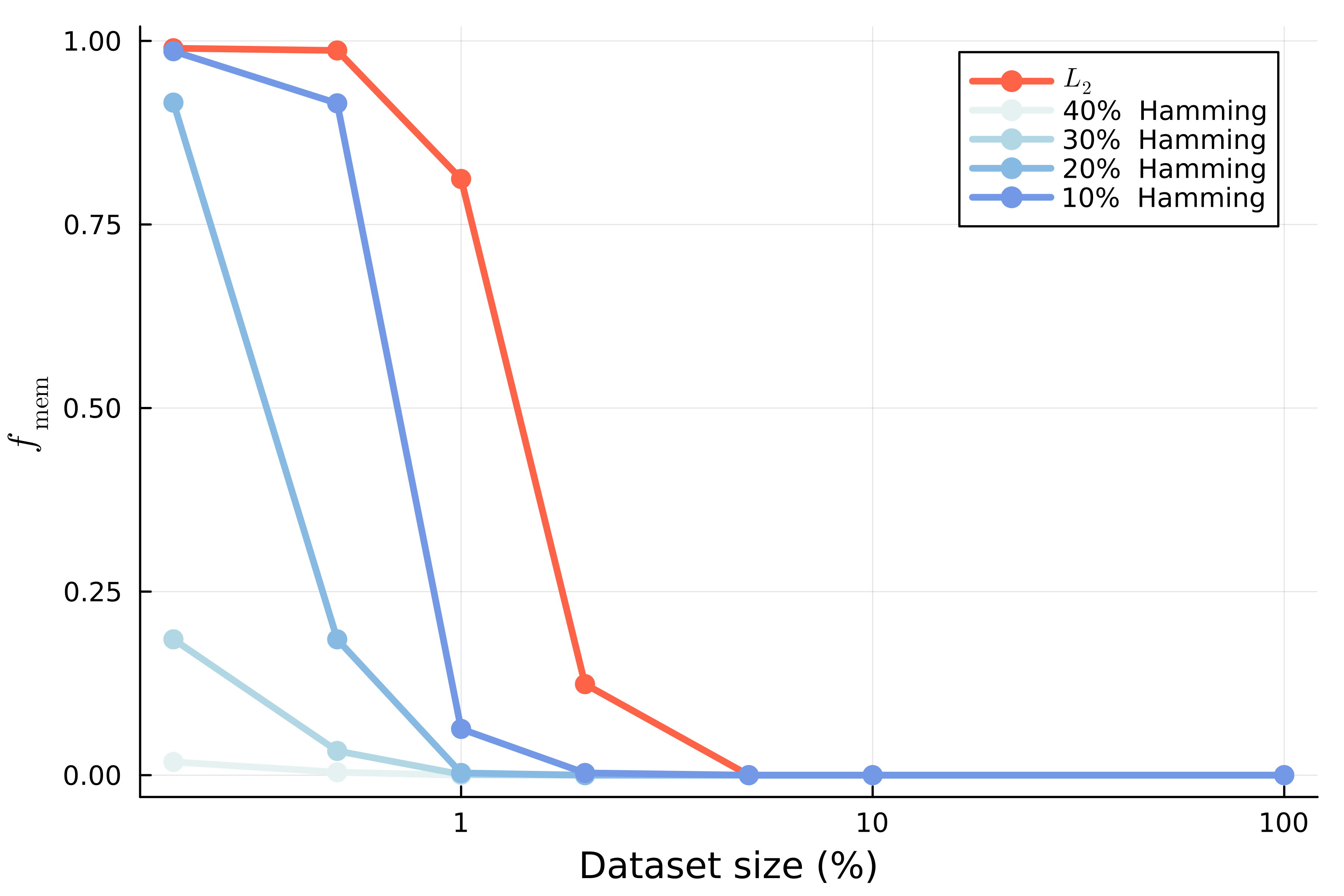}
  \caption{(Left) One minus the ratio of Eq.~\eqref{eq:ratio} as a function of the fraction $1-t$ of initially unmasked tokens, for different training set sizes ($1-t=0$ corresponds to generation from scratch).
    For each model we generated 100 images using classifier-free guidance with scale $\gamma=1$.
    (Right) Fraction of memorized images as a function of the training set size.
    The red line corresponds to the $L_2$ criterion with $\kappa=1/\sqrt{3}$.
    The blue lines correspond to the Hamming criterion: an image is considered memorized if at least the fraction of its tokens indicated in the legend coincides with that of a training image.
    For each model we generated 1000 images from scratch, using classifier-free guidance with scale $\gamma=1$ on one of the 10 classes chosen at random.
    The models are the same as in Fig.~\ref{fig:uturn-cifar}.}
  \label{fig:mem-cifar}
\end{figure}

We take $\kappa=1/\sqrt{3}$ in our experiments. Since tokens are categorical, we also consider a criterion based on the Hamming distance: an image is considered memorized if at least a given fraction of its tokens coincides with those of some training image.
As expected, for both metrics the fraction of memorized images increases as the training set size decreases, see Fig.~\ref{fig:mem-cifar}.

\section{The Linear Score Model}\label{sec:lsm}

We study a minimal score function for MDLMs, consisting of a single dense layer that we call the \textbf{Linear Score Model} (LSM).
Treating the $\mask$ token as an additional input value, the model outputs the logits of the categorical distribution over the non-mask tokens $a \in \mathcal{V}$ at each masked position $i$ as:
\begin{equation}
  p_\theta(x_i = a | x^t) = \softmax\Big(\frac{1}{\sqrt{L}} \sum_j W_{ij}^{a, x_j^t} + b_i^a\Big)
  \label{eq:lsm}
\end{equation}
where $W_{ij}^{a, x_j^t}$ is the weight connecting the input token $x_j^t \in \mathcal{V} \cup \{\mask\}$ at position $j$ to the output token $a \in \mathcal{V}$ at position $i$, and $b_i^a$ is a bias.
Note that during sampling, the strength of the categorical logits can be modified by a temperature factor $\tau^{-1}$.
In what follows, sampling is performed with $\tau=0$ which we refer to as the greedy sampling scheme.

We consider $M$ training examples, $\{\xi^\mu\}_{\mu=1}^M$, sampled from a data distribution $p_0$.
The loss of Eq.~\eqref{eq:loss} factorizes over output positions, so that the $L$ output units (the rows of $W$) can be optimized independently.

To probe memorization we perform retrieval experiments: for each training example $\xi^\mu$ in the training set and a given time $\tin$, we mask a fraction $\tin$ of the positions to obtain $x_\mu^{\tin}$. In the continuous diffusion literature, these are sometimes called U-turn experiments.
We then use $x^{\tin}$ as the starting point of the reverse process, which we run from $t=\tin$ to $t=0$, unmasking in random order to produce a final configuration $x_\mu^0$. For binary data, the retrieval quality is measured by the final magnetization averaged over the training examples:
\begin{equation}
  \mout = \frac{1}{LM} \sum_{i\mu} x^0_{i\mu}\, \xi^\mu_i
\end{equation}
which is studied theoretically in Section~\ref{sec:retrieval} for data sampled uniformly on the hypercube.

\begin{figure}[htbp]
  \centering
  \includegraphics[width=\linewidth]{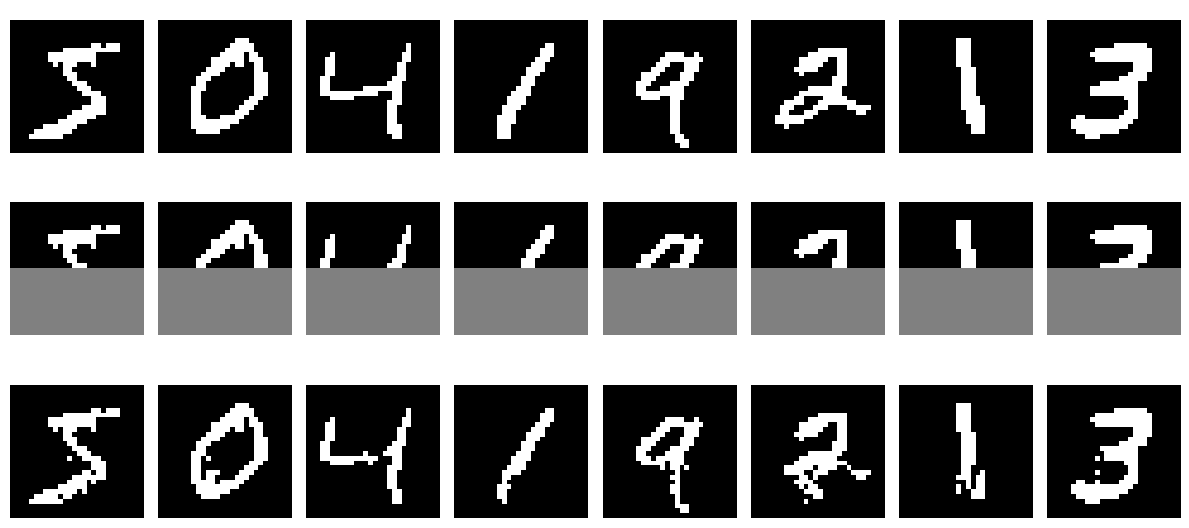}
  \caption{MNIST reconstruction with the LSM starting from the masked bottom half.
    Training set size $\alpha=0.1$ ($L=784$, $M = 78$). The model was trained with PyTorch and PyTorch Lightning, for at least 5000 epochs, with batch size 512 and the AdamW optimizer.
    (top) Original images.
    (middle) Masked images.
    (bottom) Reconstructed images after sampling one token per time step.}
  \label{fig:mnist_reconstr}
\end{figure}

To establish if the LSM is capable of meaningful memorization, we also perform experiments on real data. We train the model on binarized MNIST digits ($L=784$), occlude the bottom half of each training image and reconstruct it with the trained LSM.
As shown in Fig.~\ref{fig:mnist_reconstr}, the training digits are reconstructed with high fidelity, up to a few incorrect pixels.

\section{Replica Analysis of the LSM}\label{sec:replica}

In the case of a binary alphabet of Ising spins $x_i \in \{-1, +1\}$ we can represent the mask token with the value $0$, and the LSM of Eq.~\eqref{eq:lsm} can be rewritten as:
\begin{equation}
  p_\theta(x_i=\pm 1|x^t) = \sigma\Big(\pm \frac{1}{\sqrt{L}} \sum_{j=1}^L \big(W_{ij} x_j^t + V_{ij}\, \II(x_j^t=0)\big)\Big)
\end{equation}
where $\sigma$ is the logistic function and the corrupted sequence $x^t \in \{-1,0,+1\}^L$ is obtained by independently setting each token to $0$ with probability $t$ and leaving it unchanged otherwise.
We sample the training patterns $\mathcal{D}=\{\xi^\mu\}_{\mu=1}^M$ uniformly and independently on the hypercube $\{-1, +1\}^L$. Although the distribution has no structure to be learned, the model can still memorize the training set.
It can be shown that the symmetry imposed by the uniform sampling of the data forces $V_{ij}=0$ and we ignore these terms hereafter.
The component-wise empirical loss becomes:
\begin{equation}
  \mathcal{L}_i(w) =
  -\frac{1}{M}\sum_{\mu=1}^M
  \EE_{t \sim U(0,1)} \frac{1}{t}
  \EE_{x^t | t}\,
  \II(x_i^t=0)
  \log \sigma\Big(\frac{1}{\sqrt{L}}\, \xi_i^\mu \sum_{j} w_j x_j^t\Big)
  \label{eq:loss-LSM}
\end{equation}
where $x^t$ denotes the corrupted version of $\xi^\mu$.
Here $w$ denotes the weights of a single neuron $i$, i.e.\
$w_{j} = W_{ij}$.
The optimization is obtained independently for each row $i$ by:
\begin{equation}
  \hat{W}_i = \argmin_{w}\;\; \mathcal{L}_i(w) + \frac{1}{2}\frac{\lambda}{M}\|w\|^2.
\end{equation}

The problem is strictly convex thanks to the random masking, even for $\lambda=0$, therefore its minimizer is unique.
We are interested in the high-dimensional limit where $L, M \to \infty$ with $\alpha = M/L$ finite.

We reframe the optimization problem as a statistical mechanics problem \citep{engel2001statistical}.
We sketch the main steps leaving the details to \ref{ap:replica}.
We define the partition function
\begin{equation}
  Z_i(\beta) = \int \dif w\; e^{-\beta M \mathcal{L}_i(w) - \frac{1}{2} \beta \lambda \|w\|^2},
\end{equation}
with the goal of computing the asymptotic free entropy
\begin{equation}
  \phi(\beta) = \lim_{L \to \infty} \EE_{\mathcal{D}} \frac{1}{L} \log Z_i(\beta)
\end{equation}
in the $\beta\to\infty$ limit.
All positions $i$ are statistically equivalent.
To compute the average of the $\log$ we use the replica trick: $\EE_{\mathcal{D}} \log Z_i(\beta) = \lim_{n \to 0} \EE_{\mathcal{D}} [Z_i^n(\beta) - 1]/n.$ The expectation over the data factorises inside the replicated partition function.
Defining the $n \times n$ overlap matrix $Q_{ab} = \frac{1}{L} \sum_{j\neq i} w^a_{j} w^b_{j}$, a central limit theorem (CLT) argument on the loss term gives:
\begin{equation}
  \EE_{\mathcal{D}}\Big[e^{-\beta \sum_{a=1}^n M \mathcal{L}_{i}(w^a)}\Big] = \Big[\EE_{\{\Delta_a\}_{a=1}^n}\, e^{\beta \sum_{a=1}^n \EE_t \EE_{z_a | \Delta_a} \log \sigma(z_a)}\Big]^M
  \label{eq:CLT}
\end{equation}
where the expectations are to be taken over $z_a | \Delta_a \sim \mathcal{N}\big((1-t) \Delta_a,\, t(1-t)Q_{aa}\big)$ and $\{\Delta_a\}_{a=1}^n \sim \mathcal{N}(0,Q)$.
Enforcing the definition of the overlaps with delta functions in Fourier representation, we can write the replicated partition function as an integral over the order parameters $Q_{ab}$ and their conjugates $\hat{Q}_{ab}$.
Applying the Replica Symmetry (RS) ansatz and performing the $\beta\to\infty$ limit one obtains the free entropy as the solution of the variational problem:
\begin{equation}
  \phi = \lim_{\beta\to\infty} \frac{\phi(\beta)}{\beta} = \extr_{q,\, \delta q,\, \hat{q},\, \delta \hat{q}}\;\; -\frac{1}{2} q\, \delta \hat{q} -\frac{1}{2} \hat{q}\, \delta q + \mathcal{G}_S(\hat{q}, \delta \hat{q}) + \alpha\, \mathcal{G}_E(q, \delta q).
  \label{eq:var}
\end{equation}

\begin{figure}[htbp]
  \centering
  \includegraphics[width=0.49\linewidth]{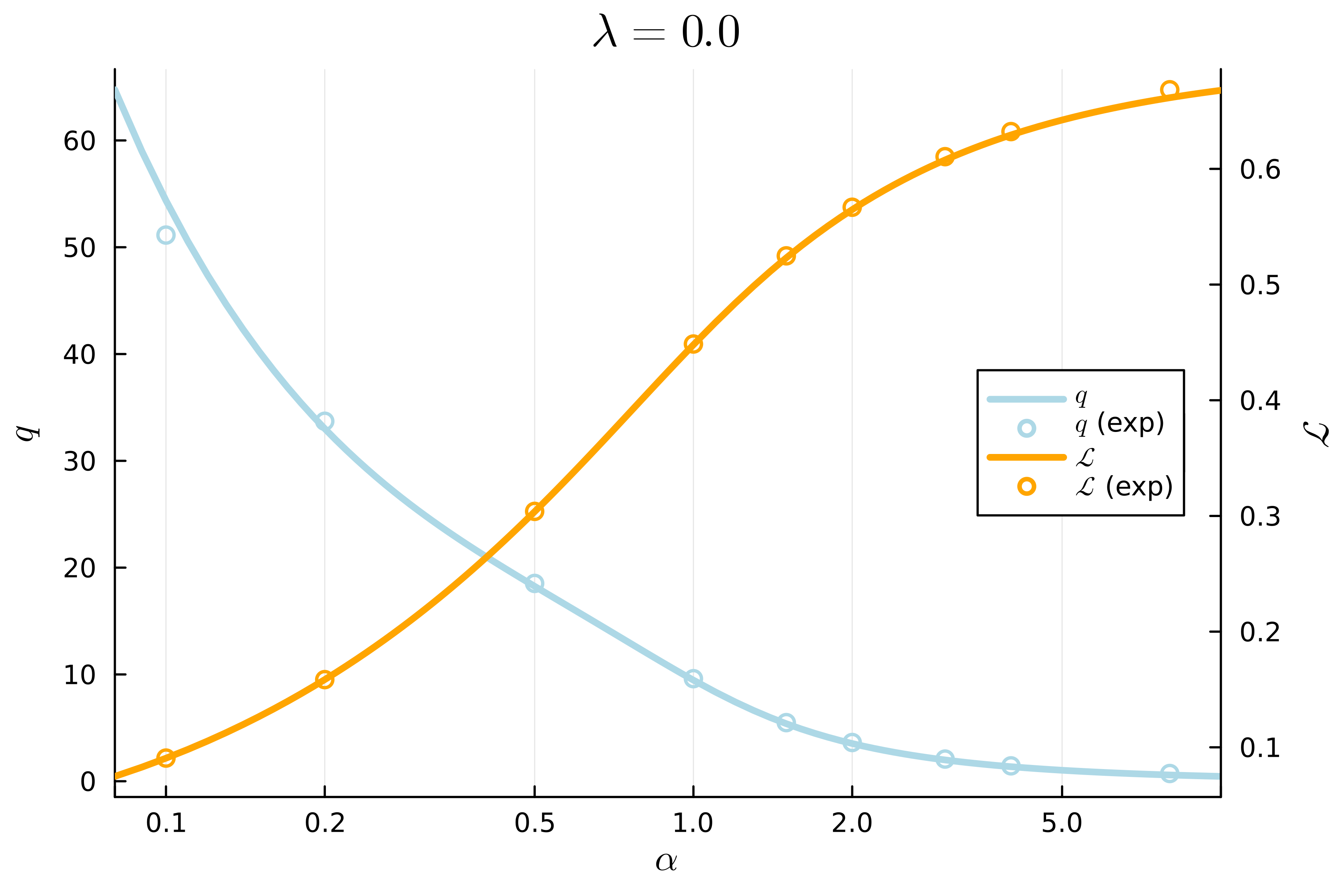}\hfill
  \includegraphics[width=0.49\linewidth]{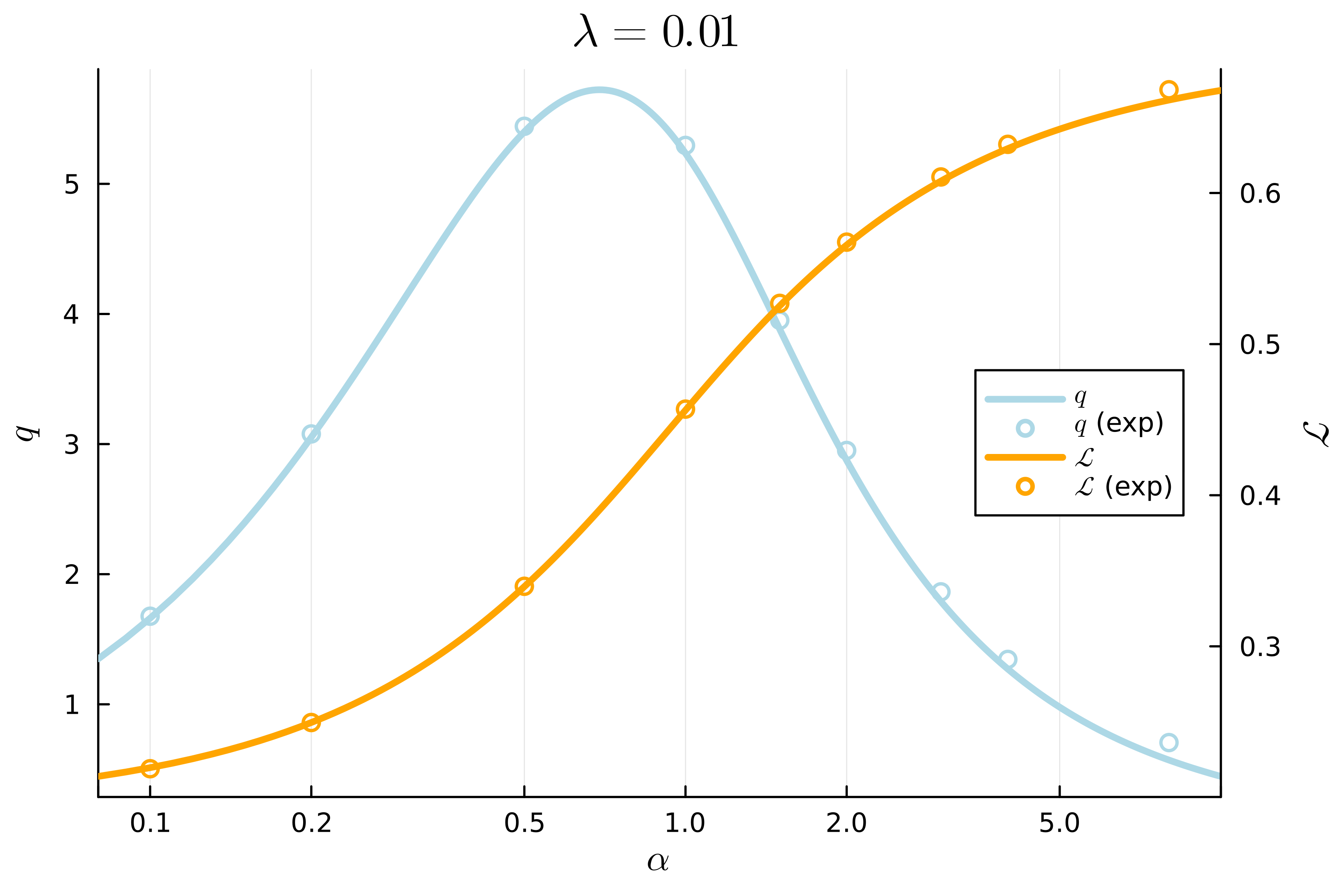}
  \caption{Squared norm $q$ of the rows of $\hat{W}$ and training loss $\mathcal{L}$ as a function of $\alpha$: saddle point predictions (lines) and trained LSM (crosses), for $\lambda=0$ (left) and $\lambda=0.01$ (right). Experimental points are row-wise averages of a single training run of the LSM on patterns of length $L=1024$ for 15000 epochs. Error bars are smaller than the markers.}
  \label{fig:SP}
\end{figure}

The entropic and energetic terms $\mathcal{G}_S$ and $\mathcal{G}_E$ and the resulting saddle point equations are given in \ref{ap:replica} (Eq.~\eqref{eq:SP}). Since the problem is convex, the RS ansatz is expected to be exact.
The RS saddle predicts the normalized square norm $q$ of each row of the learned weight matrix $\hat{W}$ and the per-position training loss $\mathcal{L} = -(\lambda q/2 + \phi)/\alpha$ (Fig.~\ref{fig:SP}).

Moreover, one can compute the stability $\hD_i^\nu = \frac{1}{\sqrt{L}}\, \xi_i^\nu \sum_j \hat{W}_{ij} \xi_j^\nu$ for each row $i$ of the learned matrix $\hat{W}$ and for some training pattern $\xi^\nu$.
It can be shown (see \ref{ap:stability}) that $\hD_i^\nu$ is a random variable whose distribution is given by the pushforward of the standard normal distribution $u \sim \mathcal{N}(0,1)$ through the solution $\hD(u)$ of the proximal map:
\begin{equation}
  \begin{aligned}
    \hD(u) & = \argmin_{\hD} \left[\frac{(\hD-\sqrt{q}\,u)^2}{2\, \delta q} + \ell(\hD,q)\right] = \prox_{\delta q\, \ell(\cdot, q)} \sqrt{q}\, u \\
    \ell(\hD,q) & = -\EE_t\, \EE_{z \sim \mathcal{N}(0,1)} \log \sigma\big(\sqrt{t(1-t)q}\, z + (1-t)\hD\big).
  \end{aligned}
  \label{eq:mu}
\end{equation}

This has to be intended in the sense that for any reasonable test function $f$, one has in distribution:
\begin{equation}
  \frac{1}{LM} \sum_{i, \mu} f(\hD_i^\mu) \to \int Du\, f(\hD(u))
  \label{eq:stability-weak-conv}
\end{equation}

\section{Retrieval Dynamics}\label{sec:retrieval}

\subsection{Accuracy}

Our goal is to understand the probability of reconstructing one of the patterns in the dataset given a masked observation of it and a network $W$.
Towards this aim, we define the accuracy $\mathcal{A}_t(c_t, e_t)$ as the probability, given some pattern $\xi^\nu$, of correctly reconstructing any token that is still masked at time $t$, given that fractions $c_t$ and $e_t$ of the tokens have been correctly and incorrectly unmasked, respectively.
Given the symmetry of the problem we expect this quantity to concentrate once averaged over different realizations, and we define it as:
\begin{equation}
  \mathcal{A}_t(c_t, e_t) := \EE_{\mathcal{D}} \EE_{C_t, E_t}\, \sigma\big(\tau^{-1} \xi_i^\nu h(C_t, E_t)\big)
  \label{eq:accuracy_def}
\end{equation}
where the preactivation $h$ is:
\begin{equation}
  h(C_t, E_t) = \frac{1}{\sqrt{L}} \sum_{j \in C_t} W_{ij} \xi_j^\nu - \frac{1}{\sqrt{L}} \sum_{j \in E_t} W_{ij} \xi_j^\nu.
\end{equation}
As we pointed out earlier, once the weights $W$ have been fixed, the sampling can be performed with different schemes depending on the value of $\tau$.
In our experiments we consider $\tau=0$, for which $\sigma(\tau^{-1} \cdot)$ becomes the Heaviside step function, often referred to as greedy decoding.
The average is performed over all uniformly random sets $C_t$ and $E_t$ of correctly and incorrectly unmasked positions at time $t$, of size $|C_t| = L c_t$ and $|E_t| = L e_t$ such that $c_t+e_t=1-t$.
We define the magnetization $m_t=c_t-e_t$, which, together with $t$, uniquely identifies $c_t$ and $e_t$. From now on we write $\mathcal{A}_t(m_t)$.

\subsection{One Step or Many Steps}

Our aim is to give a theoretical description of pattern retrieval as described in Section~\ref{sec:lsm}.
Starting at time $\tin$ with a fraction $1-\tin$ of unmasked tokens and a magnetization $\minit$ with respect to some pattern $\xi^\nu$, we run the unmasking process and obtain the final magnetization $\mout$. From now on we will always consider $\minit=c_{\mathrm{in}}=1-\tin$, meaning $e_{\mathrm{in}}=0$.

One of the distinctive features of MDLMs is that generation can be performed using any number of unmasking steps.
We focus on two limiting schedules.
In \textbf{one-step retrieval}, all remaining masked tokens are predicted simultaneously from the same initial partially observed configuration, and hence all predictions are conditioned on the same information available at the initial time.
In \textbf{step-by-step retrieval}, exactly one token is unmasked at each step, and each subsequent prediction is conditioned on the growing set of previously unmasked tokens.

In one-step retrieval, the accuracy $\mathcal{A}_{\tin}(1-\tin)$ gives an exact prediction of the expected final average magnetization.
To the initial fraction $1-\tin$ of correct tokens one must add a fraction $\tin$ with probability $\mathcal{A}_{\mathrm{in}}$ and subtract a fraction $\tin$ with probability $1-\mathcal{A}_{\mathrm{in}}$.
The final average magnetization is:
\begin{equation}
  \mout = 1 - 2\tin\big(1-\mathcal{A}_{\tin}(1-\tin)\big).
  \label{eq:onestep}
\end{equation}

Step-by-step retrieval is of course more complicated, as each newly unmasked token becomes part of the input used to predict the remaining masked ones.

One could naively attempt an analysis similar to the one-step case, where the average magnetization changes after each step of length $1/L$ as $L m_{t-1/L} = L m_t + \mathcal{A}_t(m_t) - (1-\mathcal{A}_t(m_t))$.
Taking $L\to\infty$ one would get the continuous-limit ODE for the average magnetization
\begin{equation}
  \dot{m}_t = 1-2\, \mathcal{A}_t(m_t)
  \label{eq:stepbystep-integrated-accuracy}
\end{equation}
which can be integrated numerically backwards in time, from $\tin$ to $t=0$, with the initial condition $\minit = 1-\tin$.

This approximation however fails to capture the correct dynamics, the reason being that in reality the sets $C_t$ and $E_t$ grow along a retrieval trajectory and are not uniformly sampled at each step, thus depend on previous predictions.

In what follows we analyse the dynamics for both one-step and step-by-step retrieval for a LSM with Hebbian weights and with learned weights $\hat{W}$.
In both cases the one-step dynamics is given exactly by Eq.~\eqref{eq:onestep}, while we obtain Markovian closures for the step-by-step case, exact in the Hebbian case and approximate for learned weights, where the solution is given as a system of ODEs that can be interpreted as generalizations of Eq.~\eqref{eq:stepbystep-integrated-accuracy}.

\subsection{Retrieval Dynamics with Hebbian Weights}
As a warmup, we consider the retrieval dynamics of the simpler case in which we run the decoding once the network $W$ has been prescribed Hebbian weights $H_{ij} = \frac{1}{\sqrt{L}} \sum_{\mu=1}^M \xi_i^\mu \xi_j^\mu$.
Starting from the definition of accuracy of Eq.~\eqref{eq:accuracy_def}, we can separate the signal from the noise in the preactivation:
\begin{equation}
  \xi_i^\nu h(C_t, E_t) = \frac{1}{L} \Big[ \sum_{j \in C_t} \sum_{\mu=1}^M \xi_i^\mu \xi_j^\mu \xi_i^\nu \xi_j^\nu - \sum_{j \in E_t} \sum_{\mu=1}^M \xi_i^\mu \xi_j^\mu \xi_i^\nu \xi_j^\nu \Big] \sim \mathcal{N}\big(m_t, \alpha(1-t)\big).
\end{equation}
This, in turn, gives:
\begin{equation}
  \mathcal{A}_t(m_t) = \int Dz\, \sigma\Big(\tau^{-1} \big(m_t + \sqrt{\alpha(1-t)}\,z\big)\Big).
  \label{eq:hebbs-dynamics}
\end{equation}

We plot the final magnetization $\mout$ as a function of the initial magnetization $\minit$ obtained with Eq.~\eqref{eq:onestep} for one-step retrieval for different values of $\alpha$ in Fig.~\ref{fig:mi_mf} (top left).

In this simpler Hebbian case, we show that the exact dynamics of step-by-step retrieval can be described with a Markovian closure that leads to a system of coupled ODEs for the magnetization of the retrieved pattern and the variance $v_t$ of the (rescaled) overlaps with all other patterns.
Here we provide a self-contained argument, while a derivation based on the Martin--Siggia--Rose (MSR) formalism \citep{martin1973statistical} is provided in \ref{ap:MSR}.

The magnetization relative to pattern $\xi^\nu$ can be rewritten as $m_t=\frac{1}{L} \sum_{j=1}^L \xi_j^\nu x_j^t$.
The overlaps with the other patterns are of order $1/\sqrt{L}$, and we define the rescaled overlaps $m_t^\mu=\frac{1}{\sqrt{L}} \sum_{j=1}^L \xi_j^\mu x_j^t$.
We can write the probability of the next token $x_i$ given the previously generated ones as:
\begin{equation}
  p_H(x_i = \pm 1 | x^t) =
  \sigma\big(\pm \tau^{-1} \xi_i^\nu (m_t + g_t)\big).
\end{equation}
where $i$ is the masked site selected for unmasking at time $t$ and $g_t = \frac{1}{\sqrt{L}} \sum_{\mu \neq \nu} \xi_i^\nu \xi_i^\mu m_t^\mu$. Since the entries $\xi_i^\mu$ of the still-masked site $i$ have not influenced the past trajectory, conditionally on the past $g_t$ is Gaussian with mean 0 and variance $\Var(g_t) = \alpha v_t$ where:
\begin{equation}
  v_t
  = \frac{1}{M} \sum_{\mu \neq \nu} (\xi_i^\nu \xi_i^\mu m_t^\mu)^2
  = \frac{1}{ML} \sum_{\mu \neq \nu}
  \Big(\sum_{j=1}^L \xi_j^\mu x_j^t\Big)^2.
  \label{eq:hebbian-variance}
\end{equation}

The evolution of the two variables $m_t$ and $v_t$, conditioned on the past, is:
\begin{equation}
  \begin{aligned}
    m_{t-1/L}-m_t & = \frac{\xi_i^\nu x_i}{L} \\
    v_{t-1/L}-v_t & = \frac{1}{L}
                    + \frac{2 x_i \xi_i^\nu g_t}{L \alpha}
  \end{aligned}
\end{equation}
since site $i$ is being unmasked at time $t$.
We coarse-grain the dynamics over $k$ successive unmasking steps, with $1 \ll k \ll L$, so that each interval contains many updates while $m_t$ and $v_t$ change only infinitesimally.
Fluctuations in the averaged rates then vanish as $k \to \infty$, and the Gaussian description of the crosstalk field yields closed deterministic equations for $m_t$ and $v_t$ with initial conditions $m_{\tin} = v_{\tin} = 1-\tin$:
\begin{equation}
  \begin{aligned}
    \dot{m}_t & = 1-2 \int Dz\,
               \sigma\big(\tau^{-1} (m_t + \sqrt{\alpha v_t}\, z)\big) \\
    \dot{v}_t & = -1-4 v_t \tau^{-1} \int Dz\,
               \sigma'\big(\tau^{-1} (m_t + \sqrt{\alpha v_t}\, z)\big).
  \end{aligned}
  \label{eq:hebbs-dynamics-exact}
\end{equation}

Setting $\dot{v}_t = -1$, the first equation reduces to the naive Eq.~\eqref{eq:stepbystep-integrated-accuracy} with the accuracy of Eq.~\eqref{eq:hebbs-dynamics}. Since the second term in $\dot{v}_t$ is negative we have $v_t \geq 1-t$, thus the effect of the dynamics is to increase the effective variance of the noise.
The system can be solved numerically to get the retrieval dynamics of the magnetization as a function of time and is shown in Fig.~\ref{fig:mi_mf} (bottom left).

\subsection{Retrieval Dynamics with Learned Weights}
The accuracy of Eq.~\eqref{eq:accuracy_def} can be computed for learned weights $\hat{W}$ by first conditioning on the dataset:
\begin{equation}
  \mathcal{A}_t(c_t,e_t)
  = \EE_{\mathcal{D}} \Big[
    \EE_{C_t,E_t} \big[
      \sigma\big(\tau^{-1} \xi_i^\nu h(C_t,E_t)\big) \,\big|\, \mathcal{D}
    \big]
  \Big].
  \label{eq:tower}
\end{equation}

We write
\begin{equation}
  \xi_i^\nu h(C_t,E_t)
  = \frac{1}{\sqrt{L}}\, \xi_i^\nu \sum_{j \neq i} \hat{W}_{ij} x_j^t.
\end{equation}
Recall that $x_j^t=\xi_j^\nu$ for $j \in C_t$,
$x_j^t=-\xi_j^\nu$ for $j \in E_t$, and $x_j^t=0$ otherwise.
In the large $L$ limit, conditioning on the dataset gives $\EE\, x_j^t = m_t \xi_j^\nu$ and $\Var(x_j^t)=1-t-m_t^2$.
It can be shown that the covariance for $j \neq k$ is of subleading order.
Conditioned on the patterns $\mathcal{D}$ we get:
\begin{equation}
  \begin{aligned}
    & \EE_{C_t, E_t} [ \xi_i^\nu h(C_t, E_t)] = \hD_i^\nu m_t \\
    & \Var(\xi_i^\nu h(C_t, E_t)) = q (1-t-m_t^2) .
  \end{aligned}
\end{equation}

By CLT, the inner expectation of Eq.~\eqref{eq:tower} is a function of $\hD_i^\nu$ only,
\begin{equation}
  f(\hD_i^\nu) = \int Dz\; \sigma\Big(\tau^{-1} \big(\hD_i^\nu m_t + \sqrt{q(1-t-m_t^2)}\, z\big)\Big).
\end{equation}
and the outer expectation follows from Eq.~\eqref{eq:stability-weak-conv}:
\begin{equation}
  \mathcal{A}_t(m_t) = \int Du\, Dz\, \sigma\Big(\tau^{-1} \big(\hD(u) m_t + \sqrt{q (1-t-m_t^2)}\,z\big)\Big).
  \label{eq:accuracy}
\end{equation}

\begin{figure}[htbp]
  \centering
  \includegraphics[width=\linewidth]{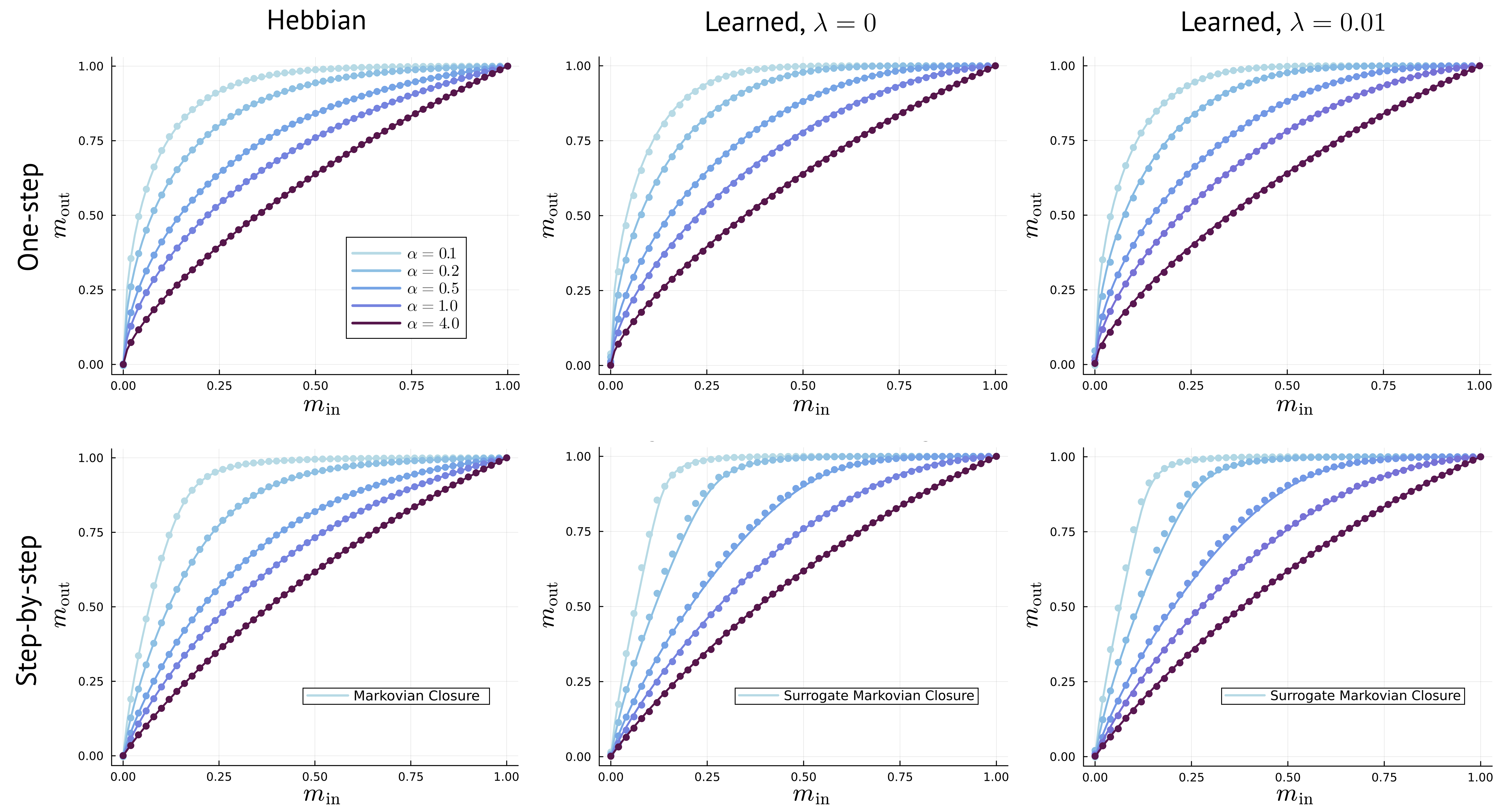}
  \caption{Final magnetization $\mout$ as a function of the initial magnetization $\minit = 1-\tin$, for Hebbian weights (left) and learned weights with $\lambda=0$ (center) and $\lambda=0.01$ (right), for several values of $\alpha$.
    Top: one-step retrieval, Eq.~\eqref{eq:onestep} (lines) and simulations (dots, average over 1000 trajectories for a single weight matrix with $L=1024$).
    Bottom: step-by-step retrieval, Markovian closure Eq.~\eqref{eq:hebbs-dynamics-exact} (left) and surrogate Markovian closure Eq.~\eqref{eq:dmft-markov} (center, right), and simulations (dots, average over 200 trajectories for a single weight matrix with $L=1024$).
    The Markovian closure is exact in the Hebbian case, while the learned-weight closure neglects two-time correlations.}
  \label{fig:mi_mf}
\end{figure}

The resulting one-step final magnetizations $\mout$ are shown in Fig.~\ref{fig:mi_mf} for $\lambda=0$ and $\lambda=0.01$ (top center and top right), in excellent agreement with simulations.

We were not able to provide an exact closure for the dynamics of the step-by-step case. Instead, we give an approximate description of it based on a Hebbian decomposition of the learned weights $\hat{W}$.
As shown in \ref{ap:hebbian-weights}, imposing stationarity of the regularized loss with respect to the weight $w_j$ gives, in the high-dimensional limit,
\begin{equation}
  \hat{W}_{ij} = \frac{1}{\sqrt{L}} \sum_{\mu=1}^M a_i^\mu \xi_i^\mu \xi_j^\mu
  \qquad\quad a_i^\mu = -\frac{\partial_{\hD} \ell(\hD_i^\mu, q)}{\lambda + \frac{2}{L} \sum_{\nu=1}^M \partial_q \ell(\hD_i^\nu, q)}
  \label{eq:hebbian-weights}
\end{equation}
where the components $a_i^\mu$ are called embedding strengths \citep{engel2001statistical}. Following the same approach of the Hebbian case, we can use this decomposition to construct a Markovian closure for this harder case.
To make things analytically tractable, we consider an approximation of the exact decomposition in which we neglect cross-pattern correlations.
We first define the iid surrogate embedding strengths as:
\begin{equation}
  \tilde{a}_i^\mu = \hD(u_i^\mu) - \sqrt{q}\, u_i^\mu
  \label{eq:surrogate-components}
\end{equation}
with $u_i^\mu \iid \mathcal{N}(0,1)$.
As shown in \ref{ap:hebbian-weights}, this choice makes the surrogate embedding strengths independent across patterns while matching the replica constraints on the normalized squared norm $q$ and on the single-pattern stability.
For the condensed pattern, the same construction identifies $u_i^\nu$ with:
\begin{equation}
  \sqrt{q}\, u_i^\nu
  = \frac{1}{\sqrt{L}} \sum_{\mu \neq \nu}
  \tilde{a}_i^\mu \xi_i^\nu \xi_i^\mu
  \Big(\frac{1}{\sqrt{L}} \sum_{j \neq i} \xi_j^\mu \xi_j^\nu\Big).
  \label{eq:surrogate-cavity}
\end{equation}

The price to pay is that this decomposition is no longer exact: independence between embedding strengths is only possible when a finite number $M$ of infinite-size patterns are stored, and indeed the original $a_i^\mu$ fall back to $\tilde{a}_i^\mu$ in the $\alpha\to 0$ limit.
Nonetheless, we will see that this approximation is able to display high levels of agreement with experimental data.

Differently from the Hebbian case, here we have that:
\begin{equation}
  p_{\hat{W}}(x_i = \pm 1 | x^t) = \sigma\Big(
    \pm \tau^{-1} \xi_i^\nu
    \big(m_t \hD(u_i^\nu) + g_t - m_t \sqrt{q}\, u_i^\nu\big)
  \Big),
\end{equation}
where now
\begin{equation}
  g_t = \frac{1}{\sqrt{L}} \sum_{\mu \neq \nu}
  \tilde{a}_i^\mu \xi_i^\nu \xi_i^\mu m_t^\mu.
\end{equation}

Since $g_t$ and $\sqrt{q}\, u_i^\nu$ are both CLT sums involving the same components $\tilde{a}_i^\mu$ they are in general correlated, and we thus look for a Gaussian decomposition which highlights this fact.
Differently from before, we have $\Var(g_t) = q v_t$ and $\Cov(g_t,\sqrt{q}\,u_i^\nu) = q r_t$ where $v_t$ is defined as in Eq.~\eqref{eq:hebbian-variance} and:
\begin{equation}
  r_t = \frac{1}{M} \sum_{\mu \neq \nu} m_t^\mu
  \Big(\frac{1}{\sqrt{L}} \sum_{j \neq i} \xi_j^\mu \xi_j^\nu\Big).
\end{equation}

Finally, we can write $g_t$ as a linear combination of two independent standard normal variables $u_i^\nu$ and $z_i^\nu$:
\begin{equation}
  g_t = r_t \sqrt{q}\, u_i^\nu
  + \sqrt{q(v_t-r_t^2)}\, z_i^\nu,
  \quad
  z_i^\nu, u_i^\nu \iid \mathcal{N}(0,1).
  \label{eq:cavity_field}
\end{equation}
We show in \ref{ap:learned-dmft} that tracking the evolution of $v_t$ and $r_t$ allows us to write the following closed system of ODEs, which we refer to as the \textbf{surrogate Markovian closure}:
\begin{equation}
  \begin{aligned}
    \dot{m}_t & = 1 - 2 \int Du\, Dz\, \sigma\big(\tau^{-1} h(u,z)\big) \\
    \dot{v}_t & = -1 - 4 \tau^{-1}\, \overline{\Delta}
               \int Du\, Dz\, \sigma'\big(\tau^{-1} h(u,z)\big)
               \Big[v_t-r_t m_t \Big(1-\frac{\hD'(u)}{\sqrt{q}}\Big)\Big] \\
    \dot{r}_t & = \dot{m}_t - 2 \tau^{-1}\, \overline{\Delta}
               \int Du\, Dz\, \sigma'\big(\tau^{-1} h(u,z)\big)
               \Big[r_t-m_t \Big(1-\frac{\hD'(u)}{\sqrt{q}}\Big)\Big].
  \end{aligned}
  \label{eq:dmft-markov}
\end{equation}
with $\overline{\Delta} = \int Du\, \hD(u)$ and where
\begin{equation}
  h(u,z) = \hD(u) m_t
  + (r_t-m_t) \sqrt{q}\, u
  + \sqrt{q(v_t-r_t^2)}\, z
\end{equation}
to be integrated backwards in time with initial conditions $m_{\tin} = v_{\tin} = r_{\tin} = 1-\tin$.
At $t=\tin$ one has $r_t = m_t$, and $h(u,z)$ reduces to the one-step field of Eq.~\eqref{eq:accuracy}. For Hebbian weights, $\tilde{a}_i^\mu = 1$ one gets $q=\alpha$ and $\hD(u) = 1 + \sqrt{\alpha}\, u$, and Eq.~\eqref{eq:dmft-markov} reduces to Eq.~\eqref{eq:hebbs-dynamics-exact}. As in the Hebbian case, the dynamics inflates the variance $v_t$ above its naive value.
Here, however, not all components of this variance are detrimental to retrieval, since the coherent part $r_t$ remains aligned with the stability $\hD(u)$.

The surrogate Markovian closure is in good agreement with simulations (Fig.~\ref{fig:mi_mf}, bottom center and bottom right).
It nevertheless neglects an order-one reaction term, which originates from the weak dependence of the previously generated spins on the couplings of the site being unmasked: unlike in the Hebbian case, the entries $\xi_i^\mu$ enter $\hat{W}$ through training. An exact treatment would therefore require two-time response functions. The resulting discrepancies are visible in Fig.~\ref{fig:trajectories}.

\subsection{Probability of Reconstructing a Full Pattern}

We consider greedy step-by-step retrieval and denote by $\Pmem$ the probability of reconstructing a full pattern.
Let $i_l$ be the position to be unmasked at step $l$ and $t_l$ the corresponding unmasking time. We have:
\begin{equation}
  \Pmem = \EE \Big[
    \prod_{l=1}^{L \tin} \Theta\big(\xi^\nu_{i_l} h(C_{t_l})\big)
  \Big]
\end{equation}
where the expectation is over the dataset $\mathcal{D}$, the initial set of correctly unmasked tokens $C_{\tin}$, and all the possible trajectories of the unmasking process. Notice that the preactivation only depends on $C_t$ since $E_t$ is empty at every step $l$.
Using $1-\prod_t \Theta(\dots) \leq \sum_t [1-\Theta(\dots)]$ we get:
\begin{equation}
  \Pmem \geq 1 - L \tin \big[1-\mathcal{A}_{\tin}(1-\tin)\big]
\end{equation}
where we used the monotonicity of $\mathcal{A}_t(1-t)$ in $t$. In \ref{ap:lb} we prove that for all $\lambda \geq 0$ and $\alpha$ small enough (for instance $\alpha\leq 1/2$) one has $\mathcal{A}_t \geq 1 - e^{-c_t/\alpha}$ for some constant $c_t>0$ depending only on $t$. We therefore have $\Pmem \geq 1 - L \tin e^{-c_{\tin}/\alpha}$, and a sufficient condition for $\Pmem\geq 1-\delta$ is:
\begin{equation}
  M \lesssim \frac{L}{\log(L/\delta)}
\end{equation}

Therefore, a number of patterns $M$ cannot be perfectly retrieved from any finite fraction of their tokens only if growing faster than $L/\log L$.

\section{Pinned Hopfield Model}\label{sec:pinned}

It is natural to ask how the standard HM behaves under a similar setup, where an initial fraction $1-t$ of its spins (here in place of tokens) is revealed and kept fixed while the remaining fraction $t$ is free to evolve.
We store the same $M = \alpha L$ independent binary patterns in a HM \citep{hopfield1982neural} with energy and couplings
\begin{equation}
  E(x) = -\frac{1}{2} \sum_{ij} J_{ij} x_i x_j, \quad
  J_{ij} = \frac{1}{L} \sum_{\mu=1}^M \xi_i^\mu \xi_j^\mu.
\end{equation}
To reconstruct a target pattern $\xi^\nu$, we reveal a uniformly random set $C$ of $(1-t)L$ positions and keep $x_i=\xi_i^\nu$ fixed on $C$ throughout the dynamics.
The remaining tokens are initialized uniformly at random and repeatedly updated by zero-temperature single-spin dynamics.
Thus $t$ is fixed during retrieval: it specifies how much of the pattern is missing, while the free spins can change their predictions as the system relaxes.
We ask how accurately they recover the missing entries as a function of $\alpha$ and $t$.

The initial magnetization is $\minit=1-t$.
Writing $m$ for the final magnetization on the free positions, that is:
\begin{equation}
  m = \frac{1}{tL} \sum_{i \notin C} \xi_i^\nu x_i
\end{equation}
the total overlap with the target is
\begin{equation}
  \mout = \frac{1}{L} \sum_i \xi_i^\nu x_i = (1-t)+t m.
\end{equation}

Within the replica-symmetric (RS) description \citep{amit1985storing}, a typical free spin sees an effective field in the target direction $\mout+\sqrt{\alpha r}\,z$, where $z\sim\mathcal{N}(0,1)$ and $\alpha r$ is the crosstalk variance, including the feedback of the free spins.
Defining the signal-to-noise ratio $y=\mout/\sqrt{\alpha r}$, the saddle point equations for the pinned HM give:
\begin{equation}
  y \sqrt{\alpha} = (1-t)+t\big[2\Phi(y)-1-2y\, \varphi(y)\big].
  \label{eq:pinned-hopfield}
\end{equation}
with total overlap $\mout=(1-t)+t[2\Phi(y)-1]$.
The derivation is given in \ref{ap:pinned-hopfield}.

\begin{figure}[htbp]
  \centering
  \includegraphics[width=0.48\linewidth]{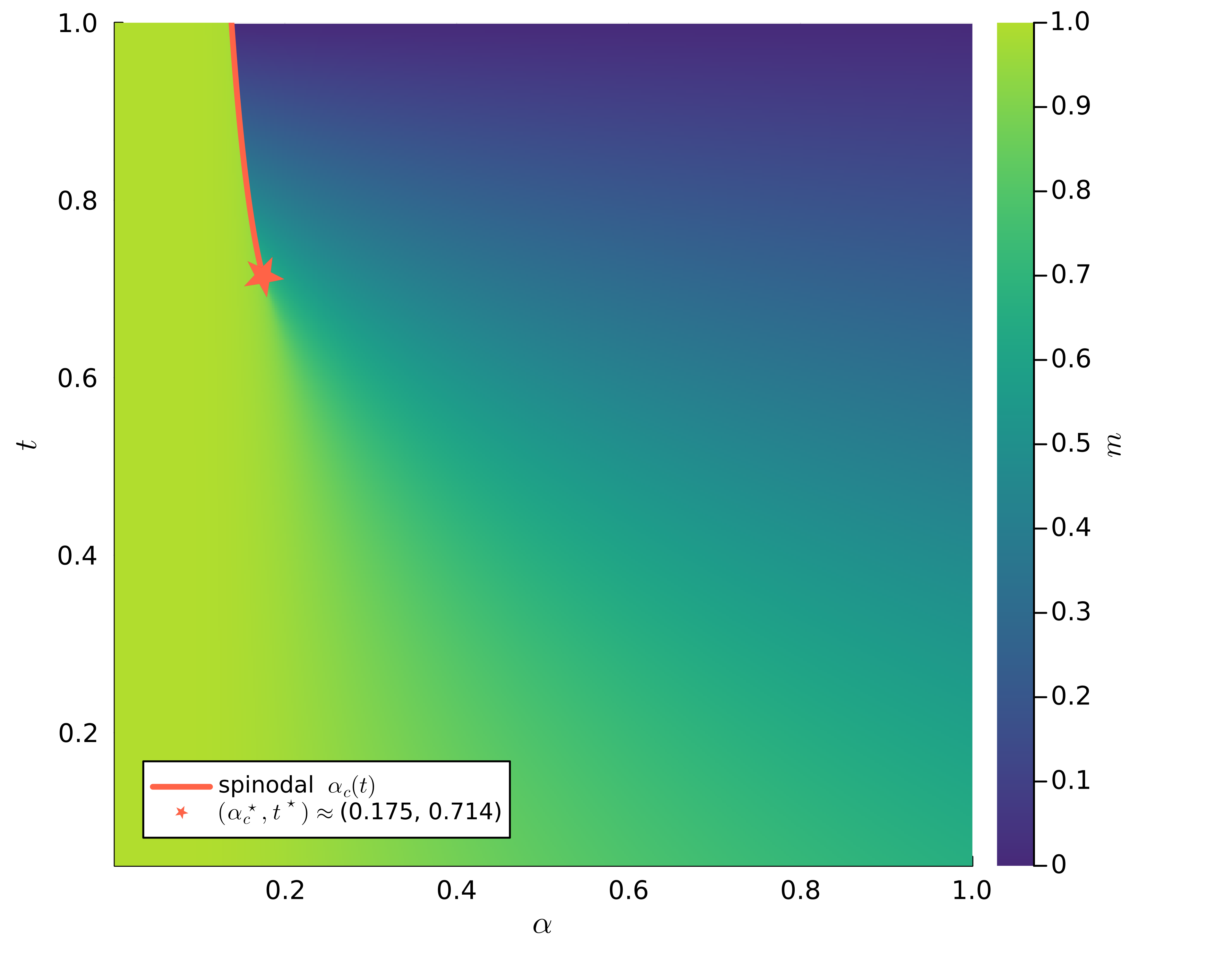}\hfill
  \includegraphics[width=0.48\linewidth]{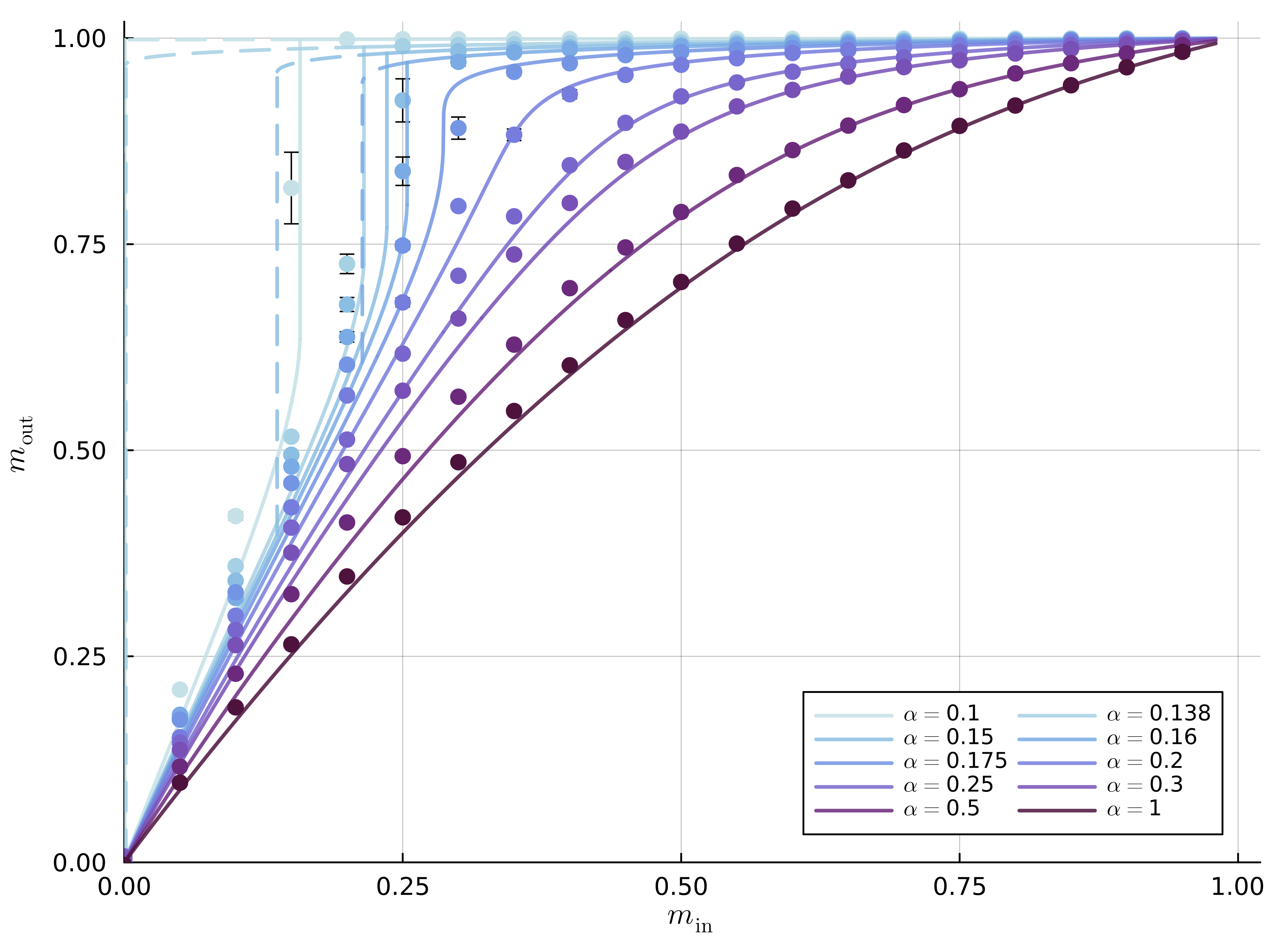}
  \caption{Pinned Hopfield model at zero temperature.
    (Left) Magnetization $m$ of the free spins in the $(\alpha, t)$ plane, from the largest root of Eq.~\eqref{eq:pinned-hopfield}. The red line is the spinodal $\alpha_c(t)$, which terminates at $(\alpha_c^*, t^*) \approx (0.175, 0.713)$.
    (Right) Final magnetization $\mout$ as a function of $\minit = 1-t$, RS solutions (lines) and zero-temperature Monte Carlo simulations with random initialization of the free spins ($L = 20000$, 10 samples). Solid lines follow the smallest root; dashed lines show the additional retrieval branch where the two coexist.}
  \label{fig:hopfield}
\end{figure}

When several positive solutions coexist, the largest root describes the retrieval branch, with the highest overlap with the target.
The phase diagram (Fig.~\ref{fig:hopfield}, left) displays a spinodal line $\alpha_c(t)$ where this branch disappears.
It grows from the Amit--Gutfreund--Sompolinsky value $\alpha_c(1) \approx 0.138$ at $t=1$ to $\alpha_c(t^*) \approx 0.175$ at $t^* \approx 0.713$, where the two spinodals meet.
For $t < t^*$, i.e.\ when more than about $29\%$ of the spins are pinned, the discontinuity disappears and the retrieval magnetization varies continuously with $\alpha$.
This phenomenology is similar to the one observed for the random pinning of glass formers \citep{cammarota2012ideal}.
In the simulations (Fig.~\ref{fig:hopfield}, right), increasing the revealed fraction $\minit$ produces a sharp improvement in retrieval at small loads, whereas the curves are smooth at larger loads.

\section{Conclusions}

We introduced the linear score model, a minimal masked diffusion model that can be analyzed exactly in the high-dimensional limit and behaves as an associative memory.
Using the replica method, we gave a precise asymptotic characterization of the trained weights, including the distribution of the pattern stabilities, from which the one-step retrieval performance follows exactly.
For step-by-step decoding, we derived an exact Markovian closure for Hebbian couplings and an approximate, but accurate, closure for the learned couplings, both in terms of low-dimensional ODE systems.

The derived $\minit$--$\mout$ phase diagram accurately captures the transition between successful and unsuccessful retrieval regimes. Finally, we showed that perfect retrieval requires a number of stored patterns sublinear in $L$.

Several directions remain open.
An exact dynamical theory for learned couplings must account for the reaction term neglected by our surrogate closure, which requires two-time response functions.
Since uniformly distributed data have no structure, our model can only memorize: extending the analysis to structured data, e.g.\ in hidden-manifold \citep{goldt2019modelling} or teacher--student settings, would allow us to describe the transition from memorization to generalization observed in practical MDLMs.
Other natural extensions include random-feature or deeper score models \citep{gerace2020generalisation,loureiro2021learning}, and the parallel decoding of several tokens per step.

\appendix
\counterwithin*{figure}{section}

\section{Replica Computation}\label{ap:replica}
\subsection{CLT on replicated Loss}
The replicated partition function reads:
\begin{equation}
  \begin{aligned}
    \EE_{\mathcal{D}} [Z_i^n(\beta)] & = \EE_{\mathcal{D}} \int \prod_a \dif w^a\,
                               e^{-\beta \sum_{a=1}^n M \mathcal{L}_i(w^a)
                               - \frac{1}{2} \beta \lambda \sum_{a=1}^n \|w^a\|^2} \\
                             & = \int \prod_a \dif w^a\,
                               \Big[\EE_\xi\, e^{-\beta \sum_{a=1}^n \ell(w^a, \xi)}\Big]^M
                               e^{-\frac{1}{2} \beta \lambda \sum_{a=1}^n \|w^a\|^2},
  \end{aligned}
\end{equation}
where we used the independence of the training patterns and defined the single-pattern loss through
\begin{equation}
  \begin{aligned}
    \sum_{a=1}^n \ell(w^a, \xi) & = -\int_0^1 \dif t\, \frac{1}{t}\,
                               \EE_{x^t | t} \Big[
                                 \II(x_i^t=0)
                                 \sum_{a=1}^n \log \sigma\Big(\frac{1}{\sqrt{L}}\, \xi_i \sum_j w_j^a x_j^t\Big)
                               \Big] \\
                             & = -\int_0^1 \dif t\,
                               \EE_{x^t | t,x_i^t=0} \Big[
                                 \sum_{a=1}^n \log \sigma\Big(\frac{1}{\sqrt{L}}\, \xi_i \sum_{j \neq i} w_j^a x_j^t\Big)
                               \Big].
  \end{aligned}
\end{equation}
The second equality follows from the fact that position $i$ is masked with probability $t$.
Conditioning on $x_i^t=0$ leaves the corruption of all other positions independent.

We claim that for fixed $t, \xi, w^a$, the variables $z_a^t= \frac{1}{\sqrt{L}}\, \xi_i \sum_{j \neq i} w_j^a x_j^t$ are asymptotically Gaussian.
Their mean is given by
\begin{equation}
  \EE_{x^t | t,x_i^t=0}\, z_a^t = (1-t)\frac{1}{\sqrt{L}}\, \xi_i \sum_{j \neq i} w_j^a \xi_j = (1-t) \Delta_a,
\end{equation}
where in the last line we defined:
\begin{equation}
  \Delta_a = \frac{1}{\sqrt{L}}\, \xi_i \sum_{j \neq i} w_j^a \xi_j.
\end{equation}
Their covariance is
\begin{equation}
  \Cov_{x^t | t,x_i^t=0}(z_a^t,z_b^t) = \big(t(1-t)\big)\frac{1}{L} \sum_{j \neq i} w_j^a w_j^b = t(1-t)\, Q_{ab},
\end{equation}
where we defined the replica overlap
\begin{equation}
  Q_{ab} = \frac{1}{L} \sum_{j \neq i} w_j^a w_j^b.
\end{equation}

Since the corruption average acts on a sum of single-replica terms, only the marginal distribution of each $z_a^t$ is needed to evaluate the single-pattern losses.
In the Gaussian approximation, this marginal is
\begin{equation}
  z_a^t \sim \mathcal{N}\big((1-t)\Delta_a,\, t(1-t)Q_{aa}\big),
\end{equation}
with $\xi$, $t$, and the weights held fixed and position $i$ masked.

At fixed replica weights, the overlaps $Q_{ab}$ do not depend on the pattern $\xi$, and their definitions can be enforced using Dirac delta functions outside the pattern average.
The quantities $\Delta_a$, however, depend on $\xi$, so their joint distribution must also be computed.
For uniformly sampled binary patterns, their moments at fixed $\xi_i$ are
\begin{equation}
  \begin{aligned}
    & \EE_{\xi | \xi_i} \Delta_a = 0, \\
    & \EE_{\xi | \xi_i} \Delta_a \Delta_b = \frac{1}{L} \sum_{j,k \neq i} w_j^a w_k^b\,
      \EE_{\xi | \xi_i} [\xi_j \xi_k] = Q_{ab}.
  \end{aligned}
\end{equation}
Under the same central limit theorem conditions, the vector
$\{\Delta_a\}_{a=1}^n$ is therefore asymptotically Gaussian with zero mean and covariance $Q$.
Putting everything together yields Eq.~\eqref{eq:CLT}.

\subsection{RS Ansatz and Saddle Point Equations}

In what follows we retain only terms of order $\beta$ because they are the only ones that will survive the zero temperature limit.
Enforcing the definition of the overlaps with delta functions in integral representation we can write the replicated partition function as
\begin{equation}
  \EE_{\mathcal{D}}\, Z_i^n(\beta) = \int \prod_{a\leq b} \dif Q_{ab}\, \dif \hat{Q}_{ab}\; e^{n L \phi_\beta(Q, \hat{Q})}.
\end{equation}
where
\begin{equation}
  \phi_\beta(Q, \hat{Q}) = - \frac{1}{2n} \sum_{a,b} Q_{ab} \hat{Q}_{ab} + G_S(\hat{Q}) + \alpha\, G_E(Q).
  \label{eq:phi_before_rs}
\end{equation}
The entropic term is:
\begin{equation}
  G_S(\hat{Q}) = \frac{1}{nL} \log \int \prod_a \dif w^a\, e^{- \frac{1}{2} \beta \lambda \sum_a \|w^a\|^2 + \frac{1}{2} \sum_{a,b} \hat{Q}_{ab}\, w^a \cdot w^b}
\end{equation}
while the energetic term is
\begin{equation}
  \begin{aligned}
    G_E(Q) & = \frac{1}{n} \log \EE_{\{\Delta_a\}_{a=1}^n \sim \mathcal{N}(0, Q)}\, e^{- \beta \sum_a \ell(\Delta_a,Q_{aa})} \\
    \ell(\Delta,q) & = -\EE_t\, \EE_{z \sim \mathcal{N}((1-t) \Delta,\, t(1-t) q)} \log \sigma(z).
  \end{aligned}
\end{equation}
We apply the following RS ansatz:
\begin{equation}
  \begin{aligned}
    Q_{aa} & = q + \delta q \hspace{1cm} & \hat{Q}_{aa} & = \hat{q} + \delta \hat{q} \\
    Q_{ab} & = q                         & \hat{Q}_{ab} & = \hat{q}.
  \end{aligned}
  \label{eq:RS}
\end{equation}

For the entropic term we have:
\begin{equation}
  \begin{aligned}
    G_S & = \frac{1}{nL} \log \int \prod_a \dif w^a\, e^{ - \frac{1}{2} \beta \lambda \sum_a \sum_i (w^a_i)^2 + \frac{1}{2} \left[\sum_i \sum_a \delta \hat{q}\, (w_i^a)^2 + \sum_i \sum_{ab} \hat{q}\, w_i^a w_i^b\right]} \\
        & = \frac{1}{nL} \log \left(\int Dy \left(\int \dif w\, e^{ - \frac{1}{2}\left[ \beta \lambda w^2 - \delta \hat{q}\, w^2 - 2 \sqrt{\hat{q}}\, w y\right]}\right)^n \right)^L \\
        & = \int Dy \log \int \dif w\, e^{ - \frac{1}{2} (\beta \lambda - \delta \hat{q}) w^2 + \sqrt{\hat{q}}\, w y} = \frac{\hat{q}}{2(\beta \lambda - \delta \hat{q})}.
  \end{aligned}
\end{equation}

We now take care of $G_E$. We compute the inverse and the determinant of $Q$ in the RS ansatz.
One has:
\begin{equation}
  (Q^{-1})_{aa} = \frac{\delta q+(n-1)q}{\delta q\, (\delta q + n q)} \quad (Q^{-1})_{ab} = \frac{-q}{\delta q\, (\delta q + n q)} \quad \det Q = \delta q^{n-1} (\delta q + n q).
\end{equation}

We get:
\begin{equation}
  \begin{aligned}
    G_E = \frac{1}{n} \log \int \prod_a \frac{\dif \Delta_a}{\sqrt{2 \pi}}\, \exp\Big\{ & - \frac{1}{2}\Big[\sum_a \Delta_a^2 \frac{1}{\delta q} - \Big(\sum_{a} \Delta_a\Big)^2 \frac{q}{\delta q\, (\delta q + n q)}\Big] \\
    & - \frac{1}{2} \log\big(\delta q^{n-1}(\delta q + n q)\big) \\
    & - \beta \sum_a \ell(\Delta_a,q+ \delta q)\Big\}.
  \end{aligned}
\end{equation}

We can perform a Hubbard--Stratonovich transformation to linearize the quadratic term in the exponent to get:
\begin{equation}
  \begin{aligned}
    G_E & = \frac{1}{n} \log \int \prod_a \frac{\dif \Delta_a}{\sqrt{2 \pi}} \int Du\, \exp\Big\{ - \frac{1}{2 \delta q} \sum_a \Delta_a^2 + \sqrt{\frac{q}{\delta q (\delta q + n q)}}\, u \sum_a \Delta_a \\
        & \qquad - \frac{1}{2} \log\big(\delta q^{n-1}(\delta q + n q)\big) - \beta \sum_a \ell(\Delta_a,q+ \delta q)\Big\} \\
        & = \frac{1}{n} \log \int Du \left( \int \frac{\dif \Delta}{\sqrt{2 \pi}}\, e^{ - \frac{1}{2 \delta q} \Delta^2 + \sqrt{\frac{q}{\delta q (\delta q + n q)}}\, u \Delta - \beta \ell(\Delta,q+ \delta q)}\right)^n \\
        & \qquad \times e^{-\frac{1}{2} \log(\delta q^{n-1}(\delta q + n q))} \\
        & = \int Du \log \int \frac{\dif \Delta}{\sqrt{2 \pi}}\, e^{ - \frac{1}{2 \delta q} \Delta^2 + \sqrt{\frac{q}{\delta q (\delta q + n q)}}\, u \Delta - \beta \ell(\Delta,q+ \delta q)} \\
        & \qquad - \frac{1}{2} \log(\delta q) - \frac{1}{2n} \log\Big(1+n \frac{q}{\delta q}\Big) \\
        & = \int Du \log \int \frac{\dif \Delta}{\sqrt{2 \pi}}\, e^{ - \frac{1}{2 \delta q} \Delta^2 + \sqrt{\frac{q}{(\delta q)^2}}\, u \Delta - \beta \ell(\Delta,q+ \delta q)} - \frac{1}{2} \log(\delta q) - \frac{1}{2} \frac{q}{\delta q}
  \end{aligned}
\end{equation}
where in the last equality we took the $n\to 0$ limit.
Finally we take care of the interaction term.
In the small $n\to 0$ limit:
\begin{equation}
  \begin{aligned}
    \phi_\beta & = - \frac{1}{2n}\, n\,(q \hat{q}+q\, \delta \hat{q} + \hat{q}\, \delta q + \delta q\, \delta \hat{q}) - \frac{1}{2n}\, n(n-1)\, q \hat{q} + G_S(\hat{q}) + \alpha\, G_E(q) \\
                  & = - \frac{1}{2} (q\, \delta \hat{q} + \hat{q}\, \delta q + \delta q\, \delta \hat{q}) + G_S(\hat{q}) + \alpha\, G_E(q).
  \end{aligned}
\end{equation}

In the zero temperature limit, we apply the scalings:
\begin{equation}
  \delta q \to \frac{\delta q}{\beta} \qquad
  \hat{q} \to \beta^2 \hat{q} \qquad
  \delta \hat{q} \to \beta\, \delta \hat{q}.
\end{equation}
We rewrite the free entropy as $\phi= \lim_{\beta\to\infty} \phi_\beta/\beta$.
We finally find Eq.~\eqref{eq:var}:
\begin{equation}
  \phi = -\frac{1}{2} q\, \delta \hat{q} -\frac{1}{2} \hat{q}\, \delta q + \mathcal{G}_S + \alpha\, \mathcal{G}_E
\end{equation}
where we defined
\begin{equation}
  \mathcal{G}_S = \lim_{\beta\to\infty} \frac{1}{\beta} G_S = \frac{1}{\beta} \left[\beta^2 \frac{\hat{q}}{2\beta (\lambda - \delta \hat{q})}\right] = \frac{\hat{q}}{2 (\lambda - \delta \hat{q})}
\end{equation}
and
\begin{equation}
  \begin{aligned}
    \mathcal{G}_E & = \lim_{\beta\to\infty} \frac{1}{\beta} G_E \\
                  & = \frac{1}{\beta} \left[\int Du \log \int \dif \Delta\, e^{ - \frac{\beta}{2 \delta q} \Delta^2 + \beta \frac{\sqrt{q}}{\delta q} u \Delta - \beta \ell\left(\Delta,q+ \frac{\delta q}{\beta}\right)} - \frac{1}{2} \log\frac{\delta q}{\beta} - \frac{1}{2} \beta \frac{q}{\delta q}\right] \\
                  & = - \int Du \left[ \frac{(\hD(u)-\sqrt{q}\,u)^2}{2\, \delta q} + \ell(\hD(u),q)\right].
  \end{aligned}
\end{equation}

Here $\hD(u)$ is the function maximising the exponent in the integral over $\Delta$:
\begin{equation}
  \hD(u) = \argmin_\Delta \left[ \frac{(\Delta-\sqrt{q}\,u)^2}{2\, \delta q} + \ell(\Delta,q)\right]
\end{equation}
Imposing stationarity for $\hat{q}$ and $\delta \hat{q}$ one gets:
\begin{equation}
  \delta \hat{q} = \lambda - \frac{1}{\delta q} \qquad \hat{q} = \frac{q}{\delta q^2}.
\end{equation}

Plugging back these expressions in $\phi$ and imposing stationarity on the remaining order parameters leads to the saddle point equations:
\begin{equation}
  \delta q = (\lambda - 2 \alpha\, \partial_q \mathcal{G}_E)^{-1} \qquad q = 2 \alpha\, \delta q^2\, \partial_{\delta q} \mathcal{G}_E.
  \label{eq:SP}
\end{equation}

The derivatives of $\mathcal{G}_E$ can be computed as
\begin{equation}
  \begin{aligned}
    \partial_q \mathcal{G}_E & = \int Du \left[\frac{u(\hD(u)-\sqrt{q}\,u)}{2\, \delta q \sqrt{q}} - \partial_q \ell(\hD(u),q)\right] \\
    \partial_{\delta q} \mathcal{G}_E & = \frac{1}{2} \int Du \left[\frac{\hD(u)- \sqrt{q}\,u}{\delta q} \right]^2
  \end{aligned}
\end{equation}
where, for our choice of activation function $\sigma$, we have
\begin{equation}
  \partial_q \ell(\hD,q) = \EE_t \EE_{z \sim \mathcal{N}(0,1)}\, \sigma\big(\sqrt{t(1-t)q}\, z+(1-t)\hD\big)\, z\, \frac{\sqrt{t(1-t)}}{2 \sqrt{q}}.
\end{equation}

The expression for the training loss can be obtained from the identity $L \phi = -M \mathcal{L} - \frac{1}{2} \lambda \|w\|^2$.

\subsection{Stability}\label{ap:stability}

Consider $\hD_i^\nu = \frac{1}{\sqrt{L}}\, \xi^\nu_i \sum_{j\neq i} \hat{W}_{ij} \xi^\nu_j$.
We write $\Delta_i^\nu(w) = \frac{1}{\sqrt{L}}\, \xi^\nu_i \sum_{j\neq i} w_j \xi^\nu_j$.
We then have
\begin{equation}
  \begin{aligned}
    \EE_{\mathcal{D}}\, f(\hD_i^\nu) & = \lim_{\beta, L \to \infty} \EE_{\mathcal{D}} \frac{1}{Z_i(\beta)} \int \dif w\, e^{- \frac{1}{2} \beta \lambda \|w\|^2} \Big[\prod_{\mu \neq \nu} e^{-\beta \ell(w, \xi^\mu)}\Big] \\
    & \quad \times e^{-\beta \ell(w, \xi^\nu)} f(\Delta_i^\nu(w)).
  \end{aligned}
\end{equation}

We write $Z_i(\beta)^{-1} = \lim_{n\to 0} Z_i(\beta)^{n-1}$:
\begin{equation}
  \begin{aligned}
    \EE_{\mathcal{D}}\, f(\hD_i^\nu) & = \lim_{\beta, L \to \infty} \lim_{n\to 0} \int \prod_a \dif w^a\, e^{- \frac{1}{2} \beta \lambda \sum_a \|w^a\|^2} \Big[\EE_\xi\, e^{-\beta \sum_a \ell(w^a, \xi)}\Big]^{M-1} \\
    & \quad \times \EE_{\xi^\nu} \Big[e^{-\beta \sum_a \ell(w^a, \xi^\nu)} f(\Delta_i^\nu(w^{a=1}))\Big] \\
    & = \lim_{\beta, L \to \infty} \lim_{n\to 0} \int \prod_a \dif w^a\, e^{- \frac{1}{2} \beta \lambda \sum_a \|w^a\|^2} \Big[\EE_\xi\, e^{-\beta \sum_a \ell(w^a, \xi)}\Big]^M \\
    & \quad \times F(\{w^a\}_{a=1}^n)
  \end{aligned}
\end{equation}
where
\begin{equation}
  F(\{w^a\}_{a=1}^n) = \frac{\EE_{\xi^\nu} e^{-\beta \sum_a \ell(w^a, \xi^\nu)} f(\Delta_i^\nu(w^{a=1}))}{\EE_{\xi^\nu} e^{-\beta \sum_a \ell(w^a, \xi^\nu)}}.
\end{equation}

We can now apply the same CLT arguments as before, getting
\begin{equation}
  F(\{w^a\}_{a=1}^n) = \frac{\EE_{\{\Delta_a\}_{a=1}^n} e^{-\beta \sum_a \ell(\Delta_a, Q_{aa})} f(\Delta_{a=1})}{\EE_{\{\Delta_a\}_{a=1}^n} e^{-\beta \sum_a \ell(\Delta_a, Q_{aa})}}
\end{equation}

One then proceeds with the same computation of the RS case to obtain
\begin{equation}
  F(q, \delta q) = \int Du\, \frac{\int \dif \Delta\, e^{-\frac{1}{2 \delta q} \Delta^2 + \sqrt{\frac{q}{\delta q^2}}\, u \Delta - \beta \ell(\Delta, q+ \delta q)} f(\Delta)}{\int \dif \Delta\, e^{-\frac{1}{2 \delta q} \Delta^2 + \sqrt{\frac{q}{\delta q^2}}\, u \Delta - \beta \ell(\Delta, q+ \delta q)}}
\end{equation}
where only the marked $\Delta^{a=1}$ survives, and we relabel it $\Delta$.
Since the remaining part of the integral is the same as the replica computation, we can write
\begin{equation}
  \EE_{\mathcal{D}}\, f(\hD_i^\nu) = \lim_{\beta, L \to \infty} \lim_{n\to 0} \int \dif q\, \dif \delta q\; e^{n L \phi_\beta(q, \delta q)} F(q, \delta q).
\end{equation}

If $F=O(1)$ it cannot shift the saddle point in the $L\to\infty$ limit, and the extremisers $q$ and $\delta q$ are those obtained from the replica computation.
Therefore we can write $\EE_{\mathcal{D}}\, f(\hD_i^\nu) = F(q, \delta q)$ where now $q$ and $\delta q$ are the solutions of the saddle point equations.
Performing the zero temperature limit imposes $F(q, \delta q) = \int Du\, f(\hD(u))$ where $\hD(u)$ is the solution of the optimization problem defined above.

One can then obtain the result stated in the Main Text by typical self averaging arguments.
Indeed one has:
\begin{equation}
  \begin{aligned}
    & \EE_{\mathcal{D}} \left|\frac{1}{LM} \sum_{i\mu} f(\hD_i^\mu) - \EE_{\mathcal{D}}\, f(\hD_i^\nu)\right|^2 \\
    & \quad \leq \frac{1}{L} \sum_i \EE_{\mathcal{D}} \left|\frac{1}{M} \sum_{\mu} f(\hD_i^\mu) - \EE_{\mathcal{D}}\, f(\hD_i^\nu)\right|^2 \\
    & \quad \leq \frac{1}{M} \Var_{\mathcal{D}}\big(f(\hD_i^\nu)\big) + \frac{M-1}{M} \Cov_{\mathcal{D}}\big(f(\hD_i^\nu), f(\hD_i^\mu)\big)
  \end{aligned}
\end{equation}
The first term goes to zero as $M\to\infty$. Moreover, carrying out the same computation of above for the product $f(\Delta_i^\nu)f(\Delta_i^\mu)$ of two distinct patterns $\nu \neq \mu$ leads to
\begin{equation}
  \EE_{\mathcal{D}}\, f(\hD_i^\nu) f(\hD_i^\mu) = \left[\int Du\, f(\hD(u))\right]^2
\end{equation}
and thus also the covariance term goes to 0.
The empirical stability distribution therefore converges weakly in probability to the predicted single-stability law.

\section{Learned Weights Hebbian Decomposition}\label{ap:hebbian-weights}

Starting from Eq.~\eqref{eq:loss-LSM} we impose stationarity with respect to the weight $w_j$.
With the same CLT argument as before, the loss of each pattern depends on $w$ only through the stability $\hD_i^\mu$ and the norm $q$. Using $\partial \hD_i^\mu / \partial w_j = \xi_i^\mu \xi_j^\mu / \sqrt{L}$ and $\partial q / \partial w_j = 2 w_j / L$ we obtain:
\begin{equation}
  \lambda\, w_j + \frac{1}{\sqrt{L}} \sum_{\mu=1}^M \partial_{\hD} \ell(\hD_i^\mu, q)\, \xi_i^\mu \xi_j^\mu + \Big(\frac{2}{L} \sum_{\mu=1}^M \partial_q \ell(\hD_i^\mu, q)\Big) w_j = 0
\end{equation}
meaning that weights belong to the space spanned by the Hebbian directions and can be written for $j \neq i$ via embedding strengths $a_i^\mu$ as:
\begin{equation}
  w_j = \frac{1}{\sqrt{L}} \sum_{\mu=1}^M a_i^\mu \xi_i^\mu \xi_j^\mu,
  \qquad
  a_i^\mu = -\frac{\partial_{\hD} \ell(\hD_i^\mu, q)}{\lambda + \frac{2}{L} \sum_{\nu=1}^M \partial_q \ell(\hD_i^\nu, q)}.
\end{equation}
In the high-dimensional limit the denominator concentrates to the same value for all rows:
\begin{equation}
  \leff = \lambda + 2 \alpha \int Du\, \partial_q \ell(\hD(u), q)
\end{equation}
which, from the saddle point equations, can be rewritten as
\begin{equation}
  \delta q\, \leff = 1 - \alpha\Big[1 - \int Du\, \frac{\hD'(u)}{\sqrt{q}}\Big].
\end{equation}

Define $u_i^\nu$ as the value that solves $\hD_i^\nu = \hD(u_i^\nu)$, which exists and is unique due to the strict monotonicity of $\hD(u)$ (\ref{ap:lb}).
The stationarity condition of the proximal problem Eq.~\eqref{eq:mu} reads $- \delta q\, \partial_{\hD} \ell(\hD(u), q) = \hD(u) - \sqrt{q}\, u$, and we can write:
\begin{equation}
  a_i^\nu = a(u_i^\nu) = \frac{\hD(u_i^\nu) - \sqrt{q}\, u_i^\nu}{\delta q\, \leff}.
\end{equation}

Notice that $a_i^\nu$ and $a_i^\mu$ are correlated.
This can be easily seen by computing
\begin{equation}
  q = \frac{1}{L} \sum_{j\neq i} w_j^2 = \frac{1}{L^2} \sum_{j\neq i} \Big(\sum_\mu a_i^\mu \xi_i^\mu \xi_j^\mu\Big)^2 = \alpha \frac{1}{M} \sum_\mu a_i^\mu \hD_i^\mu
\end{equation}
Instead, if the embedding strengths were independent of each other and of the patterns, only the terms $\mu = \nu$ would contribute to the double sum, giving $\alpha/M \sum_\mu (a_i^\mu)^2 = q(\delta q\, \leff)^{-2}$, which is inconsistent since $\delta q\, \leff < 1$ for $\alpha > 0$.

\subsection{Surrogate iid Embedding Strengths}

We now look for an approximation that allows us to have iid embedding strengths $a_i^\mu$, as it makes computations analytically tractable.

We define the surrogate iid embedding strengths $\tilde{a}_i^\mu$ satisfying by construction:
\begin{equation}
  q = \frac{1}{L} \sum_{j\neq i} w_j^2 = \frac{1}{L^2} \sum_{j\neq i} \sum_{\mu\nu} \tilde{a}_i^\mu \tilde{a}_i^\nu \xi_i^\mu \xi_i^\nu \xi_j^\nu \xi_j^\mu = \alpha \frac{1}{M} \sum_\mu (\tilde{a}_i^\mu)^2.
  \label{eq:q-surrogate}
\end{equation}

Computing the stability we get:
\begin{equation}
  \hD_i^\nu = \tilde{a}_i^\nu + \frac{1}{\sqrt{L}} \sum_{\mu\neq\nu} \tilde{a}_i^\mu \xi^\nu_i \xi_i^\mu \frac{1}{\sqrt{L}} \sum_{j\neq i} \xi_j^\mu \xi_j^\nu
  \label{eq:stability-surrogate}
\end{equation}
where by usual CLT arguments and the definition of the surrogate embedding strengths, the second term will be Gaussian with mean $0$ and variance $q$.
Writing $\tilde{a}_i^\mu=\hD_i^\mu-\sqrt{q}\,u_i^\mu$ and plugging it into Eq.~\eqref{eq:q-surrogate} yields:
\begin{equation}
  q = \alpha \frac{1}{M} \sum_\mu (\hD_i^\mu - \sqrt{q}\,u_i^\mu)^2
\end{equation}
which is familiar, as it is the empirical counterpart of the saddle point equation for $q$ (Eq.~\eqref{eq:SP}):
\begin{equation}
  q = \alpha \int Du\, (\hD(u) - \sqrt{q}\, u)^2.
\end{equation}
Therefore, for the surrogate embedding strengths to reproduce the observables of the replica computation, the value $u_i^\nu$ such that $\hD_i^\nu = \hD(u_i^\nu)$ must be exactly:
\begin{equation}
  \sqrt{q}\, u_i^\nu = \frac{1}{\sqrt{L}} \sum_{\mu\neq\nu} \tilde{a}_i^\mu \xi^\nu_i \xi_i^\mu \frac{1}{\sqrt{L}} \sum_{j\neq i} \xi_j^\mu \xi_j^\nu.
\end{equation}

This interpretation however cannot hold in the proportional limit for every pattern $\nu$ while simultaneously requiring all embedding strengths $\tilde{a}_i^\mu$ to be independent.
This is consistent with the fact that the surrogate decomposition is only exact in the $\alpha\to 0$ limit, where a finite number of patterns is stored.

We can compute the error introduced on each embedding strength $a_i^\nu$ by replacing it with $\tilde{a}_i^\nu$:
\begin{equation}
  \frac{a_i^\nu-\tilde{a}_i^\nu}{a_i^\nu}
  =
  \alpha
  \Big[
    1-
    \int Du\,
    \frac{\hD'(u)}{\sqrt{q}}
  \Big].
\end{equation}

Convexity of $\ell$ implies that the argument in square brackets belongs to $[0,1]$ and thus the relative correction is at most of order $\alpha$.

\section{Surrogate Markovian Closure}\label{ap:learned-dmft}

We finish the derivation of Eq.~\eqref{eq:dmft-markov} for the iid surrogate embedding strengths.
The increment of the variance is:
\begin{equation}
  v_{t-1/L}-v_t
  =
  \frac{1}{L}
  +
  \frac{2 \Gamma_i \xi_i^\nu x_i}{\alpha L},
  \qquad
  \Gamma_i
  =
  \frac{1}{\sqrt{L}}
  \sum_{\mu \neq \nu}
  \xi_i^\nu \xi_i^\mu m_t^\mu.
\end{equation}

Notice that in the Hebbian case we could directly close the dynamics with 2 ODEs since $\Gamma_i$ was replaced directly with $g_t$.
Instead the evolution of the coherent part is:
\begin{equation}
  r_{t-1/L}-r_t
  =
  \frac{\xi_i^\nu x_i}{L}
  \Big(
    1 + \frac{\Xi_i}{\alpha}
  \Big),
  \qquad
  \Xi_i
  :=
  \frac{1}{L}
  \sum_{\mu \neq \nu}
  \xi_i^\nu \xi_i^\mu
  \sum_{j \neq i}
  \xi_j^\mu \xi_j^\nu.
  \label{eq:learned-exact-increments}
\end{equation}

We thus have, implicitly conditioning on the past:
\begin{equation}
  \begin{aligned}
    \dot{m}_t & = -\EE[\xi_i^\nu x_i], \\
    \dot{v}_t & = -1 - \frac{2}{\alpha} \EE[\xi_i^\nu x_i \Gamma_i], \\
    \dot{r}_t & = \dot{m}_t - \frac{1}{\alpha} \EE[\xi_i^\nu x_i \Xi_i].
  \end{aligned}
  \label{eq:learned-macroscopic-increments}
\end{equation}

The variables $\Gamma_i,
\Xi_i,
u_i^\nu,
z_i^\nu$
are jointly Gaussian, and from the independence of the surrogate embedding strengths across patterns we have:
\begin{equation}
  \begin{aligned}
    \Cov(\Gamma_i,\sqrt{q}\,u_i^\nu) & = \alpha\, \overline{\Delta}\, r_t,
                                    & \qquad
                                    \Cov(\Gamma_i,g_t)
                                    & = \alpha\, \overline{\Delta}\, v_t, \\
       \Cov(\Xi_i,\sqrt{q}\,u_i^\nu) & = \alpha\, \overline{\Delta},
                                    &
                                    \Cov(\Xi_i,g_t)
                                    & = \alpha\, \overline{\Delta}\, r_t.
  \end{aligned}
  \label{eq:surrogate-response-covariances}
\end{equation}

We then consider Eq.~\eqref{eq:cavity_field}.
Inverting it we have:
\begin{equation}
  z_i^\nu
  =
  \frac{g_t-r_t \sqrt{q}\,u_i^\nu}{\sqrt{q(v_t-r_t^2)}}.
\end{equation}

It follows that
\begin{equation}
  \Cov(\Gamma_i,z_i^\nu) =
  \frac{\alpha\, \overline{\Delta}\, v_t - r_t\, \alpha\, \overline{\Delta}\, r_t}{\sqrt{q(v_t-r_t^2)}} =
  \alpha\, \overline{\Delta}
  \sqrt{\frac{v_t-r_t^2}{q}},
\end{equation}
whereas $\Cov(\Xi_i,z_i^\nu)=0$.
Intuitively, $\Xi_i$ probes only the coherent direction associated with the cavity variable $u_i^\nu$, while $\Gamma_i$ also contains a component along the direction $z_i^\nu$.

The conditional mean of the generated spin is
\begin{equation}
  \EE[\xi_i^\nu x_i | u_i^\nu,z_i^\nu]
  =
  2\sigma\big(\tau^{-1} h(u_i^\nu,z_i^\nu)\big)-1.
\end{equation}
Since $\Gamma_i$ and $\Xi_i$ are centered,
\begin{equation}
  \begin{aligned}
    \EE[\xi_i^\nu x_i \Gamma_i] & =
                              2\EE\big[
                                \Gamma_i\,
                                \sigma\big(\tau^{-1} h(u_i^\nu,z_i^\nu)\big)
                              \big], \\
       \EE[\xi_i^\nu x_i \Xi_i] & =
                              2\EE\big[
                                \Xi_i\,
                                \sigma\big(\tau^{-1} h(u_i^\nu,z_i^\nu)\big)
                              \big].
  \end{aligned}
  \label{eq:learned-spin-response}
\end{equation}

We first compute the response involving $\Gamma_i$.
Gaussian integration by parts gives
\begin{equation}
  \EE\big[
    \Gamma_i\,
    \sigma(\tau^{-1} h)
  \big] =
      \Cov(\Gamma_i,u_i^\nu)\,
      \EE\big[
        \partial_{u_i^\nu}
        \sigma(\tau^{-1} h)
      \big] +
      \Cov(\Gamma_i,z_i^\nu)\,
      \EE\big[
        \partial_{z_i^\nu}
        \sigma(\tau^{-1} h)
      \big].
  \label{eq:tau-Gaussian-ibp}
\end{equation}

The two derivatives of the field $h(u,z)$ are:
\begin{equation}
  \partial_u h(u,z)
  =
  m_t \hD'(u)
  +
  (r_t-m_t)\sqrt{q},
  \qquad\quad
  \partial_z h(u,z)
  =
  \sqrt{q(v_t-r_t^2)}.
\end{equation}

Therefore,
\begin{equation}
  \EE\big[
    \Gamma_i\,
    \sigma(\tau^{-1} h)
  \big]
  =
  \tau^{-1} \alpha\, \overline{\Delta}
  \int Du\, Dz\,
  \sigma'\big(\tau^{-1} h(u,z)\big)
  \Big[
    v_t
    -
    r_t m_t
    \Big(
      1-\frac{\hD'(u)}{\sqrt{q}}
    \Big)
  \Big].
\end{equation}

The computation involving $\Xi_i$ is simpler because its covariance with $z_i^\nu$ vanishes.
Gaussian integration by parts gives
\begin{equation}
  \EE\big[
    \Xi_i\,
    \sigma(\tau^{-1} h)
  \big]
  =
  \Cov(\Xi_i,u_i^\nu)\,
  \EE\big[
    \partial_{u_i^\nu}
    \sigma(\tau^{-1} h)
  \big],
\end{equation}
and hence
\begin{equation}
  \EE\big[
    \Xi_i\,
    \sigma(\tau^{-1} h)
  \big]
  =
  \tau^{-1} \alpha\, \overline{\Delta}
  \int Du\, Dz\,
  \sigma'\big(\tau^{-1} h(u,z)\big)
  \Big[
    r_t
    -
    m_t
    \Big(
      1-\frac{\hD'(u)}{\sqrt{q}}
    \Big)
  \Big].
\end{equation}
Putting everything together gives Eq.~\eqref{eq:dmft-markov}.

\begin{figure}[htbp]
  \centering
  \includegraphics[width=\linewidth]{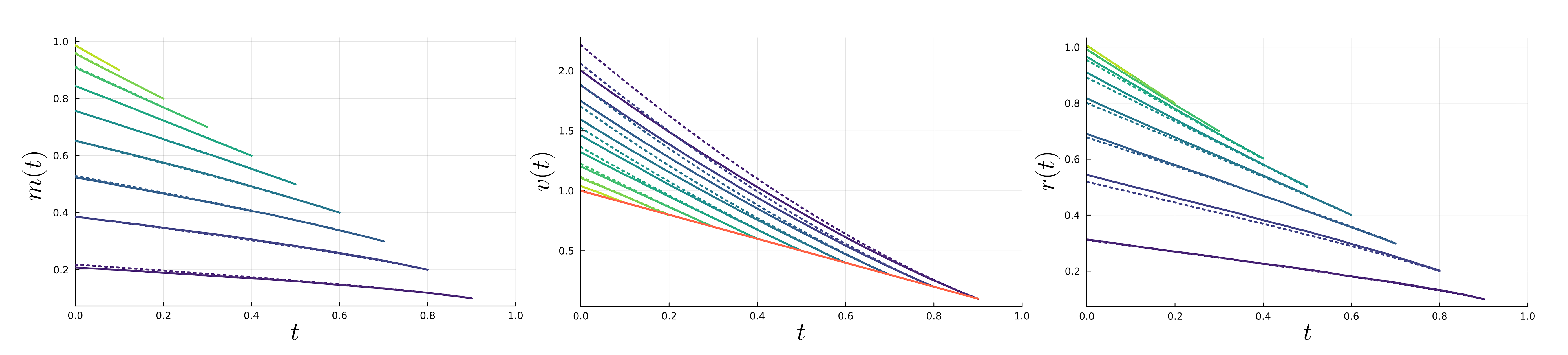}
  \caption{Trajectories of the three variables $m_t$, $v_t$, $r_t$ (left to right) of the surrogate Markovian closure from initial time $\tin$ to $t=0$ at $\alpha=1$ and $\lambda=0$, for $\minit = 1-\tin$ and $\tin = 0.1, \dots, 0.9$.
    Solid lines are experimental values averaged over $200$ runs, while dotted lines are the solution of the ODEs Eq.~\eqref{eq:dmft-markov}.
    In the plot for $v_t$ the solution corresponding to the naive $v_t = 1-t$ is shown in red.}
  \label{fig:trajectories}
\end{figure}

Sample trajectories are shown in Fig.~\ref{fig:trajectories}.
While the surrogate Markovian closure captures the dynamics of the magnetization with good accuracy, a closer inspection shows small discrepancies in the other two variables. Comparison with the naive solution of Eq.~\eqref{eq:stepbystep-integrated-accuracy}, for which $v_t = 1-t$, shows that the surrogate Markovian closure is a significant improvement.

\section{Greedy Accuracy Lower Bound}\label{ap:lb}

We focus on the greedy algorithm.
Throughout, we use the fact that $\partial_{\hD} \ell(\hD, q)\leq 0$ and $\partial_{\hD \hD} \ell(\hD, q)\geq 0$.
From Eq.~\eqref{eq:mu} we find the minimizer via the implicit equation $- \hD(u) + \sqrt{q}\,u - \delta q\, \partial_{\hD} \ell(\hD(u),q)=0$.
Define $\hD_0=\hD(0)$: we have $\hD_0 = - \delta q\, \partial_{\hD} \ell(\hD_0,q) \geq 0$.
Differentiating the implicit equation on both sides we have:
\begin{equation}
  \hD'(u) = \frac{\sqrt{q}}{1+\delta q\, \partial_{\hD \hD} \ell(\hD, q)}
\end{equation}
and thus $\hD(u)$ is nondecreasing.
For greedy decoding and $m_t = 1-t$, Eq.~\eqref{eq:accuracy} gives $1-\mathcal{A}_t = \int Du\, \Phi\big(-\hD(u) \sqrt{(1-t)/(q t)}\big)$, where $\Phi$ is the standard normal cumulative distribution function.
We split the integration of $\mathcal{A}_t$ around some $u^*$ such that $\hD(u^*)=\hD_0/2$:
\begin{equation}
  1- \mathcal{A}_t = \int_{u<u^*} Du\, \Phi\Big(-\frac{\hD(u)}{\sqrt{q}} \sqrt{\frac{1-t}{t}}\Big) + \int_{u\geq u^*} Du\, \Phi\Big(-\frac{\hD(u)}{\sqrt{q}} \sqrt{\frac{1-t}{t}}\Big).
\end{equation}

We have that $u^*$ satisfies
\begin{equation}
  u^* = \frac{\hD_0}{\sqrt{q}} \left[\frac{1}{2} - \frac{\partial_{\hD} \ell(\hD_0/2, q)}{\partial_{\hD} \ell(\hD_0, q)}\right]\leq -\frac{\hD_0}{2\sqrt{q}}
\end{equation}
because $\partial_{\hD} \ell(\hD, q)$ is nondecreasing in $\hD$.
For the $u<u^*$ region we use the trivial bound:
\begin{equation}
  \int_{u<u^*} Du\, \Phi\Big(-\frac{\hD(u)}{\sqrt{q}} \sqrt{\frac{1-t}{t}}\Big) \leq \PP(u<u^*) \leq \Phi\Big(-\frac{\hD_0}{2\sqrt{q}}\Big).
\end{equation}

In the $u\geq u^*$ region we use the bound $\hD(u) \geq \hD_0/2$ to get
\begin{equation}
  \int_{u\geq u^*} Du\, \Phi\Big(-\frac{\hD(u)}{\sqrt{q}} \sqrt{\frac{1-t}{t}}\Big) \leq \Phi\Big(-\frac{\hD_0}{2\sqrt{q}} \sqrt{\frac{1-t}{t}}\Big)
\end{equation}
so finally, for some constant $c_t$ that depends only on $t$,
\begin{equation}
  1-\mathcal{A}_t \leq \Phi\Big(-\frac{\hD_0}{2\sqrt{q}}\Big) + \Phi\Big(-\frac{\hD_0}{2\sqrt{q}} \sqrt{\frac{1-t}{t}}\Big)\leq \exp\Big(-c_t \frac{\hD_0^2}{q}\Big).
\end{equation}

Let's now take Eq.~\eqref{eq:SP}.
The saddle point for $q$ gives:
\begin{equation}
  \frac{q}{\alpha} = \int Du\, (\hD(u) - u \sqrt{q})^2 = \int Du\, \tilde{a}(u)^2
\end{equation}
where we used the definition $\tilde{a}(u) = \hD(u) - u \sqrt{q}$ of the previous section as a shorthand.
Moreover $\tilde{a}(0) = \hD_0$ and $\tilde{a}'(u) = \hD'(u)-\sqrt{q}\leq 0$, so it is non-increasing in $u$.
We split the RHS integral into the region $u\geq 0$ and $u<0$.
We have
\begin{equation}
  \int_{u\geq 0} Du\, \tilde{a}(u)^2 \leq \frac{\hD_0^2}{2}
\end{equation}
and, using the monotonicity of $\hD(u)$ and the fact that in this region $u = -|u|$:
\begin{equation}
  \int_{u<0} Du\, \tilde{a}(u)^2 \leq \int_{u<0} Du\, (\hD_0+|u| \sqrt{q})^2.
\end{equation}
Using $(a+b)^2\leq 2a^2+2b^2$ gives
\begin{equation}
  \int_{u<0} Du\, \tilde{a}(u)^2 \leq \hD_0^2+ 2 q \int_{u<0} Du\, u^2 = \hD_0^2+ q.
\end{equation}
We have that the bound
\begin{equation}
  \frac{q}{\alpha} \leq \frac{3}{2} \hD_0^2 + q \rightarrow \frac{\hD_0^2}{q} \geq \frac{2}{3} \Big(\frac{1}{\alpha} -1\Big) \geq \frac{1}{k \alpha}
\end{equation}
holds for some positive $k$ provided $\alpha \leq (1-\frac{3}{2k})$.
In this regime, for $\alpha$ small enough, we have
\begin{equation}
  1-\mathcal{A}_t \leq \exp\Big(-\frac{c_t}{\alpha}\Big).
\end{equation}
where we absorbed all other constants in $c_t$.

\section{Martin--Siggia--Rose Approach for Autoregressive Hebbian Retrieval}\label{ap:MSR}

We derive here the dynamical mean-field equations used in the Hebbian retrieval section, using the Martin--Siggia--Rose formalism \citep{martin1973statistical}.
We fix the autoregressive order as follows: at position $i$ the generated spin is conditioned on the already-unmasked spins $j>i$.
With the continuous time $t=i/L$, the reverse diffusion dynamics is therefore integrated from larger $t$ to smaller $t$.

We take Hebbian weights $H_{ij} = \frac{1}{\sqrt{L}} \sum_{\mu=1}^M \xi_i^\mu \xi_j^\mu$
and write the trajectory probability as
\begin{equation}
  p_X(x)
  =
  \prod_{i=1}^L
  \sigma\Big(
    \tau^{-1} x_i \frac{1}{L}
    \sum_{\mu=1}^M \xi_i^\mu
    \sum_{j>i} \xi_j^\mu x_j
  \Big).
\end{equation}
We condition on the retrieval of pattern $\xi^{\mu=1}$ and gauge the spins as $s_i=\xi_i^{\mu=1} x_i$.
For $\mu>1$ we absorb the condensed pattern into the disorder, $\xi_i^\mu \to \xi_i^{\mu=1} \xi_i^\mu$, which leaves the distribution iid Rademacher.
The macroscopic overlap and weak pattern overlaps are then
\begin{equation}
  m_i = \frac{1}{L} \sum_{j>i} s_j,
  \quad
  m_i^\mu = \frac{1}{\sqrt{L}} \sum_{j>i} \xi_j^\mu s_j.
\end{equation}
The signed local field is
\begin{equation}
  h_i
  =
  s_i \Big(
    m_i
    + \frac{1}{\sqrt{L}} \sum_{\mu>1} \xi_i^\mu m_i^\mu
  \Big),
\end{equation}
so that
\begin{equation}
  p_X(s | \xi^{\mu=1})
  =
  \int \prod_i \dif h_i
  \prod_i \Big[
    \sigma(\tau^{-1} h_i)\,
    \delta\Big(
      h_i - s_i \Big(
        m_i
        + \frac{1}{\sqrt{L}} \sum_{\mu>1} \xi_i^\mu m_i^\mu
      \Big)
    \Big)
  \Big].
\end{equation}

The Martin--Siggia--Rose (MSR) generating functional
\begin{equation}
  Z_L[m]
  =
  \EE_{\mathcal{D}} \sum_s
  p_X(s | \xi^{\mu=1})
  \prod_i \delta\Big(m_{i-1}-m_i-\frac{s_i}{L}\Big)
\end{equation}
constrained on a macroscopic trajectory $m$ is then (from now on sums over $\mu$ run across $\mu>1$):
\begin{equation}
  \begin{aligned}
    Z_L[m] & =
             \EE_{\mathcal{D}} \sum_s
             \int \prod_i \dif h_i
             \prod_{i\mu} \dif m_{i-1}^\mu
             \prod_i \sigma(\tau^{-1} h_i)
             \times \\
           & \quad \times
             \prod_i \delta\Big(
               h_i - s_i \Big(
                 m_i + \frac{1}{\sqrt{L}} \sum_\mu \xi_i^\mu m_i^\mu
               \Big)
             \Big)
             \times \\
           & \quad \times
             \prod_{\mu i} \delta\Big(m_{i-1}^\mu-m_i^\mu-\frac{\xi_i^\mu s_i}{\sqrt{L}}\Big)
             \prod_i \delta\Big(m_{i-1}-m_i-\frac{s_i}{L}\Big) \\
           & =
             \EE_{\mathcal{D}} \sum_s
             \int
             \prod_i \frac{L\, \dif \hat{m}_i}{2 \pi}
             \prod_i \frac{\dif h_i\, \dif \hat{h}_i}{2 \pi}
             \prod_{i\mu}
             \frac{\dif m_{i-1}^\mu\, \dif \hat{m}_i^\mu}{2 \pi}\,
             e^{\sum_i \log \sigma(\tau^{-1} h_i)}
             \times \\
           & \quad \times
             e^{
             -i \sum_i \hat{h}_i \left[
               h_i - s_i \left(
                 m_i + \frac{1}{\sqrt{L}} \sum_\mu \xi_i^\mu m_i^\mu
               \right)
             \right]
             }
             \times \\
           & \quad \times
             e^{
             -i \sum_{\mu i} \hat{m}_i^\mu \left[
               m_{i-1}^\mu - m_i^\mu
               - \frac{1}{\sqrt{L}} \xi_i^\mu s_i
             \right]
             - i L \sum_i \hat{m}_i \left[
               m_{i-1}-m_i-\frac{s_i}{L}
             \right]
             }, \\
           & \quad m_L^\mu=0, \quad m_L=0
  \end{aligned}
\end{equation}
where we used the Fourier representation of the delta functions defining $h_i$, $m_i^\mu$ and $m_i$.
The coefficients in the exponent are chosen so that all terms are order $L$.

\subsection{Disorder Average}

The disorder average factorises as
\begin{equation}
  \EE_{\xi_i^\mu}
  \exp\Big(
    i \xi_i^\mu s_i\,
    \frac{\hat{h}_i m_i^\mu + \hat{m}_i^\mu}{\sqrt{L}}
  \Big)
  =
  \cos\Big(\frac{\hat{h}_i m_i^\mu + \hat{m}_i^\mu}{\sqrt{L}}\Big).
\end{equation}
In the high-dimensional limit, we expand
$\log \cos\big(\frac{\theta}{\sqrt{L}}\big) = -\frac{\theta^2}{2L}+O\big(\frac{1}{L^2}\big)$:
\begin{equation}
  \sum_{i\mu}
  \log \cos\Big(\frac{\hat{h}_i m_i^\mu + \hat{m}_i^\mu}{\sqrt{L}}\Big)
  =
  -\frac{1}{2L} \sum_{i\mu} \big[
    (\hat{h}_i m_i^\mu)^2
    + 2\hat{h}_i m_i^\mu \hat{m}_i^\mu
    + (\hat{m}_i^\mu)^2
  \big]
  + o(L).
\end{equation}

We can compute the following Gaussian integral:
\begin{equation}
  \begin{aligned}
    & \int \frac{\dif \hat{m}_i^\mu}{\sqrt{2 \pi}}\,
      e^{
      -\frac{1}{2L} \left[
        (\hat{h}_i m_i^\mu)^2
        + 2\hat{h}_i m_i^\mu \hat{m}_i^\mu
        + (\hat{m}_i^\mu)^2
      \right]
      - i \hat{m}_i^\mu (m_{i-1}^\mu-m_i^\mu)
      } \\
    & =
      e^{-\frac{1}{2L} (\hat{h}_i m_i^\mu)^2}
      \int \frac{\dif \hat{m}_i^\mu}{\sqrt{2 \pi}}\,
      e^{
      -\frac{1}{2L} (\hat{m}_i^\mu)^2
      - \hat{m}_i^\mu \left[
        \frac{1}{L} \hat{h}_i m_i^\mu
        + i(m_{i-1}^\mu-m_i^\mu)
      \right]
      } \\
    & =
      e^{-\frac{1}{2L} (\hat{h}_i m_i^\mu)^2}
      \sqrt{L}\,
      e^{
      \frac{L}{2} \left(
        \frac{1}{L} \hat{h}_i m_i^\mu
        + i(m_{i-1}^\mu-m_i^\mu)
      \right)^2
      } \\
    & =
      \sqrt{L}\,
      e^{
      i \hat{h}_i m_i^\mu (m_{i-1}^\mu-m_i^\mu)
      - \frac{L}{2} (m_{i-1}^\mu-m_i^\mu)^2
      }.
  \end{aligned}
\end{equation}

We can then integrate over $m_i^\mu$.
This time we cannot factorise it completely and have to keep the product over $i$:
\begin{equation}
  \begin{aligned}
    & L^{L/2}
      \int \prod_{i=1}^L
      \frac{\dif m_{i-1}^\mu}{\sqrt{2 \pi}}\,
      e^{
      \sum_{i=1}^L
      i \hat{h}_i m_i^\mu (m_{i-1}^\mu-m_i^\mu)
      - \frac{L}{2} \sum_{i=1}^L (m_{i-1}^\mu-m_i^\mu)^2
      } \\
    & =
      \int \prod_{i=1}^L
      \frac{\dif m_{i-1}^\mu}{\sqrt{2 \pi}}\,
      e^{
      \sum_{i=1}^L
      i \frac{\hat{h}_i}{L} m_i^\mu (m_{i-1}^\mu-m_i^\mu)
      - \frac{1}{2} \sum_{i=1}^L (m_{i-1}^\mu-m_i^\mu)^2
      },
  \end{aligned}
\end{equation}
where we rescaled $m_i^\mu \to m_i^\mu/\sqrt{L}$.
We could compute this Gaussian integral as the determinant of a (tridiagonal) matrix.
Alternatively, here we compute it recursively by integrating one variable at a time.
Suppose at one point we have integrated all $m_k^\mu$ for $k>i$: this will have produced a new Gaussian term that only depends on $m_i^\mu$.
With the initial condition $q_L=1$ and $D_L=1$ and $m_L^\mu=0$ we have:
\begin{equation}
  \begin{aligned}
    & \int \prod_{k=i}^{L-1}
      \frac{\dif m_k^\mu}{\sqrt{2 \pi}}\,
      e^{
      \sum_{k=i}^L \left[
        i \frac{\hat{h}_k}{L} m_k^\mu (m_{k-1}^\mu-m_k^\mu)
        - \frac{1}{2} (m_{k-1}^\mu-m_k^\mu)^2
      \right]
      } \\
    & =
      \frac{1}{\sqrt{D_i q_i}}\,
      e^{-\frac{(m_{i-1}^\mu)^2}{2q_i}}.
  \end{aligned}
\end{equation}

Then inductively:
\begin{equation}
  \begin{aligned}
    & \frac{1}{\sqrt{D_i q_i}}\,
      e^{-\frac{(m_{i-1}^\mu)^2}{2q_i}} \\
    & =
      \int \prod_{k=i}^{L-1}
      \frac{\dif m_k^\mu}{\sqrt{2 \pi}}\,
      e^{
      \sum_{k=i}^L \left[
        i \frac{\hat{h}_k}{L} m_k^\mu (m_{k-1}^\mu-m_k^\mu)
        - \frac{1}{2} (m_{k-1}^\mu-m_k^\mu)^2
      \right]
      } \\
    & =
      \frac{1}{\sqrt{D_{i+1} q_{i+1}}}
      \int \frac{\dif m_i^\mu}{\sqrt{2 \pi}}\,
      e^{
      i \frac{\hat{h}_i}{L} m_i^\mu (m_{i-1}^\mu-m_i^\mu)
      - \frac{1}{2} (m_{i-1}^\mu-m_i^\mu)^2
      - \frac{(m_i^\mu)^2}{2q_{i+1}}
      } \\
    & =
      \frac{1}{\sqrt{D_{i+1} q_{i+1}}}\,
      e^{-\frac{1}{2} (m_{i-1}^\mu)^2}
      \int \frac{\dif m_i^\mu}{\sqrt{2 \pi}}\,
      e^{
      \left(1+i \frac{\hat{h}_i}{L}\right) m_{i-1}^\mu m_i^\mu
      - \left(
        \frac{1+q_{i+1}}{2q_{i+1}}
        + i \frac{\hat{h}_i}{L}
      \right)(m_i^\mu)^2
      } \\
    & =
      \frac{1}{\sqrt{
        D_{i+1}
        \left(1+q_{i+1}+2i \hat{h}_i q_{i+1}/L\right)
      }}\,
      e^{
      -\frac{1+\hat{h}_i^2 q_{i+1}/L^2}{2\left(1+q_{i+1}+2i \hat{h}_i q_{i+1}/L\right)}
      (m_{i-1}^\mu)^2
      }.
  \end{aligned}
\end{equation}
from which we get the recursion
\begin{equation}
  q_i
  =
  \frac{1+q_{i+1}+2i \hat{h}_i q_{i+1}/L}{1+\hat{h}_i^2 q_{i+1}/L^2}
  =
  q_{i+1}
  +
  \frac{(1+i \hat{h}_i q_{i+1}/L)^2}{1+\hat{h}_i^2 q_{i+1}/L^2}
\end{equation}
and
\begin{equation}
  D_i
  =
  D_{i+1}
  \big(1+\hat{h}_i^2 q_{i+1}/L^2\big).
\end{equation}
\begin{equation}
  \int \frac{\dif m_0^\mu}{\sqrt{2 \pi}}\,
  \frac{1}{\sqrt{D_1 q_1}}\,
  e^{-\frac{(m_0^\mu)^2}{2q_1}}
  =
  \frac{1}{\sqrt{D_1}}.
\end{equation}

To obtain a compact notation we go to the continuous limit, interpreting the time $t$ as $t=i/L$.
Since $q_i$ is extensive we set $q_i=L\, v(i/L)$.
Taking $L\to\infty$ yields
\begin{equation}
  \dot{v}(t) = -\big(1+i \hat{h}(t) v(t)\big)^2
\end{equation}
with $v(1)=0$ while
\begin{equation}
  \log D_1
  =
  \sum_{i=1}^{L-1} \log\frac{D_i}{D_{i+1}}
  =
  \int_0^1 \dif t\, \hat{h}(t)^2 v(t)
  + o(1).
\end{equation}

Collecting the condensed spin constraint, the weak-pattern determinant, and the determinant recursion enforced by a Lagrange multiplier $\eta(t)$, the continuum MSR action can be written as
\begin{equation}
  Z_L[m]
  =
  \int
  \mathcal{D}h\, \mathcal{D}\hat{h}\, \mathcal{D}\hat{m}\, \mathcal{D}\eta\, \mathcal{D}v\;
  e^{L \mathcal{S}[h,\hat{h},m,\hat{m},\eta,v]},
\end{equation}
with $\mathcal{S}=\int_0^1 \dif t\, S(t)$ and:
\begin{equation}
  S =
  \log \sigma(\tau^{-1} h)
  + \log\big(2 \cosh(i \hat{h} m+i \hat{m})\big)
  + i \hat{m} \dot{m}
  - i \hat{h} h
  - \frac{\alpha}{2} \hat{h}^2 v
  + \eta \big[\dot{v}+(1+i \hat{h} v)^2\big].
\end{equation}
In formulas here and below we drop the explicit time dependence $t$ in the notation.
We compute the last Gaussian integrals in $h$ and $\hat{h}$.
Expanding the square
$(1+i \hat{h} v)^2=1+2i \hat{h} v-\hat{h}^2 v^2$
and writing:
\begin{equation}
  2\cosh(i \hat{h} m+i \hat{m})
  =
  \sum_{p=\pm 1}
  e^{i p \hat{h} m+i p \hat{m}}
\end{equation}
we get, for each $p=\pm 1$:
\begin{equation}
  \begin{aligned}
    & \int \frac{\dif \hat{h}\, \dif h}{2 \pi}\,
      e^{
      -\frac{1}{2} (\alpha v+2\eta v^2)\hat{h}^2
      - i(h-2\eta v)\hat{h}
      + i p \hat{h} m + i p \hat{m}
      + \log \sigma(\tau^{-1} h)
      } \\
    & =
      e^{i p \hat{m}}
      \int
      \frac{\dif h}{\sqrt{2 \pi (\alpha v+2\eta v^2)}}\,
      \sigma(\tau^{-1} h)\,
      e^{
      -\frac{(h-2\eta v-p m)^2}{2(\alpha v+2\eta v^2)}
      } \\
    & =
      e^{i p \hat{m}}
      \int Dz\,
      \sigma\Big(
        \tau^{-1} \big(
          p m+2\eta v
          +\sqrt{\alpha v+2\eta v^2}\, z
        \big)
      \Big),
  \end{aligned}
\end{equation}
which leads to the final (local) action:
\begin{equation}
  \begin{aligned}
    S = {} &
    i \hat{m} \dot{m}
    + \eta(\dot{v}+1) \\
    & + \log\Big(
      \sum_{p=\pm 1}
      e^{i p \hat{m}}
      \int Dz\,
      \sigma\Big(
        \tau^{-1} \big(
          p m+2\eta v
          +\sqrt{\alpha v+2\eta v^2}\, z
        \big)
      \Big)
    \Big).
  \end{aligned}
  \label{eq:action}
\end{equation}

\subsection{Functional Saddle Point}

Our goal is now to find functions $v(t)$, $\eta(t)$, $\hat{m}(t)$ and especially $m(t)$ that maximise the action.
Define:
\begin{equation}
  G_p(m,\hat{m},v,\eta)
  =
  e^{i p \hat{m}}
  \int Dz\,
  \sigma\Big(
    \tau^{-1} \big(
      p m+2\eta v
      +\sqrt{\alpha v+2\eta v^2}\, z
    \big)
  \Big).
\end{equation}

Thanks to the symmetry $\sigma(-x) = 1-\sigma(x)$, the Euler--Lagrange equations obtained by varying $m$ and $v$ admit the symmetric solution $\hat{m}=0$ and $\eta=0$.
Indeed, with $G_{\pm}$ defined as above, one has:
\begin{equation}
  \begin{aligned}
              G_+(m,0,v,0) & = 1-G_-(m,0,v,0), \\
    \partial_m G_+(m,0,v,0) & = -\partial_m G_-(m,0,v,0), \\
    \partial_v G_+(m,0,v,0) & = -\partial_v G_-(m,0,v,0).
  \end{aligned}
\end{equation}

Then the Euler--Lagrange equations on $m$ and $v$ give:
\begin{equation}
  i \dot{\hat{m}}
  =
  \frac{\partial_m G_+(m,\hat{m},v,\eta) + \partial_m G_-(m,\hat{m},v,\eta)}{G_+(m,\hat{m},v,\eta) + G_-(m,\hat{m},v,\eta)}
\end{equation}
and
\begin{equation}
  \dot{\eta}
  =
  \frac{\partial_v G_+(m,\hat{m},v,\eta) + \partial_v G_-(m,\hat{m},v,\eta)}{G_+(m,\hat{m},v,\eta) + G_-(m,\hat{m},v,\eta)},
\end{equation}
which are satisfied by $\hat{m}=0$ and $\eta=0$.
Then varying the action with respect to $\hat{m}$ at the symmetric saddle gives
\begin{equation}
  i \dot{m}
  + \partial_{\hat{m}} \log(G_+ + G_-)
  = 0,
\end{equation}
hence
\begin{equation}
  \dot{m}
  =
  1-2 \int Dz\,
  \sigma\big(\tau^{-1} (m+\sqrt{\alpha v}\, z)\big).
\end{equation}
Varying with respect to $\eta$ gives
\begin{equation}
  \dot{v}+1+\partial_\eta \log(G_+ + G_-)=0.
\end{equation}
Since
\begin{equation}
  \partial_\eta \big(
    p m+2\eta v+\sqrt{\alpha v+2\eta v^2}\, z
  \big)\Big|_{\eta=0}
  =
  2v+\frac{v^2}{\sqrt{\alpha v}}\, z,
\end{equation}
the $z$-odd contributions cancel between $p=+$ and $p=-$, while the remaining terms give
\begin{equation}
  \partial_\eta (G_+ + G_-)\Big|_{\hat{m}=0,\eta=0}
  =
  4 \tau^{-1} v \int Dz\,
  \sigma'\big(\tau^{-1} (m+\sqrt{\alpha v}\, z)\big).
\end{equation}
Putting everything together we obtain Eq.~\eqref{eq:hebbs-dynamics-exact}. Here we considered generation from the fully masked sequence, $m(1) = v(1) = 0$. Retrieval from a partially observed pattern corresponds to integrating the same equations from $\tin$, with the initial conditions $m(\tin) = v(\tin) = 1-\tin$.

\section{Pinned Hopfield Model}\label{ap:pinned-hopfield}

\subsection{Replica-Symmetric Equations}

We derive the stationary equations used in Section~\ref{sec:pinned}, taking
$L,M\to\infty$ at fixed $\alpha=M/L$ and $0<t\leq 1$.
Let $C$ be the set of $(1-t)L$ pinned positions, with $x_i=\xi_i^\nu$ for $i \in C$.
Omitting the constant from the diagonal couplings, the conditional partition function is
\begin{equation}
  Z(\beta) = \sum_{\{x_{\setminus C}\}}
  e^{\frac{\beta}{2L} \sum_{\mu=1}^M \left(\sum_i \xi_i^\mu x_i\right)^2},
\end{equation}
where $\{x_{\setminus C}\}$ denotes a sum over the binary spins outside $C$.
We apply the replica trick as in \ref{ap:replica} and average over the non-target patterns, keeping $\xi^\nu$ fixed.
All replicated sums $\{x^a_{\setminus C}\}$ below obey the same pinning constraint.
\begin{equation}
  \EE_{\mathcal{D} | \xi^\nu} [Z(\beta)^n] = \sum_{\{x^a_{\setminus C}\}}
  e^{\frac{\beta}{2L} \sum_a \left(\sum_i \xi_i^\nu x_i^a\right)^2}
  \prod_{\mu \neq \nu} \EE_{\xi^\mu}\,
  e^{\frac{\beta}{2L} \sum_a \left(\sum_i \xi_i^\mu x_i^a\right)^2}.
\end{equation}
For each non-target pattern and replica we introduce an independent Gaussian variable $z_\mu^a$ with a Hubbard--Stratonovich transformation. The average over the disorder then gives:
\begin{equation}
  \begin{aligned}
    \EE_{\mathcal{D} | \xi^\nu} [Z(\beta)^n] & = \sum_{\{x^a_{\setminus C}\}}
                                      e^{\frac{\beta}{2L} \sum_a \left(\sum_i \xi_i^\nu x_i^a\right)^2}
                                      \prod_{\mu \neq \nu} \int \prod_a Dz_\mu^a\,
                                      \EE_{\xi^\mu}\, e^{\sqrt{\beta/L} \sum_i \xi_i^\mu \sum_a z_\mu^a x_i^a} \\
                                    & = \sum_{\{x^a_{\setminus C}\}}
                                      e^{\frac{\beta}{2L} \sum_a \left(\sum_i \xi_i^\nu x_i^a\right)^2}
                                      \prod_{\mu \neq \nu} \int \prod_a Dz_\mu^a
                                      \prod_i \cosh\Big(\sqrt{\beta/L} \sum_a z_\mu^a x_i^a\Big) \\
                                    & = \sum_{\{x^a_{\setminus C}\}}
                                      e^{\frac{\beta}{2L} \sum_a \left(\sum_i \xi_i^\nu x_i^a\right)^2}
                                      \Big[\int \prod_a Dz_a\,
                                        e^{\sum_i \log \cosh\left(\sqrt{\beta/L} \sum_a z_a x_i^a\right)}\Big]^{M-1},
  \end{aligned}
\end{equation}

We define the magnetization and overlap matrix on the free spins as
\begin{equation}
  m_a = \frac{1}{tL} \sum_{i \notin C} \xi_i^\nu x_i^a, \quad
  Q_{ab} = \frac{1}{tL} \sum_{i \notin C} x_i^a x_i^b.
\end{equation}
Since the pinned spins are fixed to $\xi^\nu$ in every replica, we recover the usual order parameters of the HM as:
\begin{equation}
  \frac{1}{L} \sum_i \xi_i^\nu x_i^a = (1-t)+t m_a, \quad
  \frac{1}{L} \sum_i x_i^a x_i^b = (1-t)+t Q_{ab}.
\end{equation}
Expanding $\log\cosh u=u^2/2$ in the high-dimensional limit, the Gaussian integral gives:
\begin{equation}
  \begin{aligned}
    & \log \int \prod_a \frac{\dif z_a}{\sqrt{2 \pi}}\,
    e^{-\frac{1}{2} \sum_{ab} z_a \left[\delta_{ab}-\beta\left((1-t)+t Q_{ab}\right)\right] z_b} \\
    & \quad = -\frac{1}{2} \log \det\big(I-\beta((1-t)\mathbf{1}\mathbf{1}^\top+t Q)\big).
  \end{aligned}
\end{equation}

We enforce the definitions of $m_a$ and $Q_{ab}$ with conjugate variables $\hat{m}_a$ and $\hat{Q}_{ab}$ for $a<b$, with $Q_{aa}=1$.
The full replicated partition function then reads:
\begin{equation}
  \begin{aligned}
    \EE_{\mathcal{D} | \xi^\nu} [Z(\beta)^n] & = \int \prod_a \frac{\dif m_a\, \dif \hat{m}_a}{2 \pi}
    \prod_{a<b} \frac{\dif Q_{ab}\, \dif \hat{Q}_{ab}}{2 \pi} \\
    & \quad \times \exp\Big\{\frac{\beta L}{2} \sum_a ((1-t)+t m_a)^2
    - t L \sum_a \hat{m}_a m_a \\
    & \qquad - t L \sum_{a<b} \hat{Q}_{ab} Q_{ab} - \frac{\alpha L}{2} \log \det\big(I-\beta((1-t)\mathbf{1}\mathbf{1}^\top+t Q)\big)\Big\} \\
    & \quad \times \prod_{i \notin C} \sum_{\{x_i^a=\pm 1\}}
    e^{\xi_i^\nu \sum_a \hat{m}_a x_i^a + \sum_{a<b} \hat{Q}_{ab} x_i^a x_i^b} \\
    & = \int \prod_a \dif m_a\, \dif \hat{m}_a
    \prod_{a<b} \dif Q_{ab}\, \dif \hat{Q}_{ab}\;
    e^{n L (G_I+\alpha G_E+G_S)}.
  \end{aligned}
\end{equation}
We apply the RS ansatz:
\begin{equation}
  m_a=m, \quad \hat{m}_a=\hat{m}, \quad
  Q_{aa}=1, \quad Q_{ab}=q, \quad \hat{Q}_{ab}=\hat{q}.
\end{equation}
The determinant becomes
$(1-\beta t(1-q))I-\beta((1-t)+t q)\mathbf{1}\mathbf{1}^\top$ thus:
\begin{equation}
  G_E = -\frac{1}{2} \log\big(1-\beta t(1-q)\big)
  + \frac{\beta[(1-t)+t q]}{2[1-\beta t(1-q)]}.
\end{equation}

The target pattern and the constraints contribute
\begin{equation}
  \begin{aligned}
    G_I & = \lim_{n\to 0} \Big[\frac{\beta}{2n} \sum_a ((1-t)+t m_a)^2
            - \frac{t}{n} \sum_a \hat{m}_a m_a - \frac{t}{n} \sum_{a<b} \hat{Q}_{ab} Q_{ab}\Big] \\
        & = \frac{\beta}{2} ((1-t)+t m)^2 - t \hat{m} m + \frac{t}{2} \hat{q} q.
  \end{aligned}
\end{equation}
For a free site with target entry $\xi^\nu$, the spin trace is
\begin{equation}
  \begin{aligned}
    G_S & = t \lim_{n\to 0} \frac{1}{n} \log \sum_{\{x^a=\pm 1\}}
          e^{\hat{m} \xi^\nu \sum_a x^a + \frac{\hat{q}}{2}\left[\left(\sum_a x^a\right)^2-n\right]} \\
        & = -t \frac{\hat{q}}{2} + t \lim_{n\to 0} \frac{1}{n} \log \int Dz
          \sum_{\{x^a=\pm 1\}} e^{(\hat{m} \xi^\nu+\sqrt{\hat{q}}\, z)\sum_a x^a} \\
        & = -t \frac{\hat{q}}{2} + t \int Dz \log\big[2 \cosh(\hat{m}+\sqrt{\hat{q}}\, z)\big]
  \end{aligned}
\end{equation}

The free entropy per spin is the stationary value of $\phi=G_I+\alpha G_E+G_S$.
Stationarity in $m$ and $q$ gives:
\begin{equation}
  \hat{m}=\beta((1-t)+t m), \quad
  \hat{q}=\alpha \beta^2 \frac{(1-t)+t q}{(1-\beta t(1-q))^2},
\end{equation}
while stationarity in the conjugates yields, after writing $\hat{q}=\alpha \beta^2 r$:
\begin{equation}
  \begin{aligned}
    m & = \int Dz \tanh\big(\beta((1-t)+t m+\sqrt{\alpha r}\, z)\big), \\
    q & = \int Dz \tanh^2\big(\beta((1-t)+t m+\sqrt{\alpha r}\, z)\big), \\
    r & = \frac{(1-t)+t q}{(1-\beta t(1-q))^2}.
  \end{aligned}
  \label{eq:pinned-sp}
\end{equation}

As a sanity check, at $t=1$ these reduce to the usual Hopfield RS equations.

\subsection{Zero Temperature Limit}

We take $\beta\to\infty$ with $\beta(1-q)=\delta q$.
In this limit the noise variance $r$ becomes
\begin{equation}
  r = \frac{1}{(1-t\, \delta q)^2}.
  \label{eq:pinned-r}
\end{equation}

Let $\Phi$ and $\varphi$ denote the standard normal cumulative distribution function and density, respectively.
The magnetization equation becomes
\begin{equation}
  m = \int Dz\, \sign\big((1-t)+t m+\sqrt{\alpha r}\, z\big) = 2\Phi\Big(\frac{(1-t)+t m}{\sqrt{\alpha r}}\Big)-1.
\end{equation}
To determine $\delta q$, we perform the change of variables $(1-t)+t m+\sqrt{\alpha r}\, z\to z$ in the equation for $1-q$:
\begin{equation}
  \delta q=\beta(1-q) = \int \frac{\dif z}{\sqrt{2 \pi \alpha r}}\,
  e^{-\frac{(z-((1-t)+t m))^2}{2 \alpha r}}\, \beta \sech^2(\beta z) = \frac{2\varphi\left(\frac{(1-t)+t m}{\sqrt{\alpha r}}\right)}{\sqrt{\alpha r}},
\end{equation}
where $\beta \sech^2(\beta h)\to 2\delta(h)$.
Using Eq.~\eqref{eq:pinned-r} gives $\sqrt{\alpha r} = \sqrt{\alpha}+t\, \delta q \sqrt{\alpha r}$.
Defining the signal-to-noise ratio $y=((1-t)+t m)/\sqrt{\alpha r}$ we get:
\begin{equation}
  (1-t)+t[2\Phi(y)-1] = y \sqrt{\alpha}+2t y\, \varphi(y)
\end{equation}
which, rearranging, is Eq.~\eqref{eq:pinned-hopfield}.

\subsection{Retrieval Spinodal}
Write $g(y)=2\Phi(y)-1-2y\, \varphi(y)$.
The spinodal is the last point where there are two roots of Eq.~\eqref{eq:pinned-hopfield}. We impose that the derivative of the left-hand side with respect to $y$ is equal to the derivative of the right-hand side, which gives
$\sqrt{\alpha}=t\, 2y^2 \varphi(y)$.
Eliminating $\alpha$ gives
\begin{equation}
  \frac{1-t}{t}
  = 2y(y^2+1)\varphi(y)-2\Phi(y)+1.
  \label{eq:pinned-spinodal}
\end{equation}
The derivative of the right-hand side with respect to $y$ is $2y^2(2-y^2)\varphi(y)$, so the spinodal condition for $t$ occurs at $y=\sqrt{2}$.
The two spinodals therefore meet at
\begin{equation}
  t^* = \big[2+6\sqrt{2}\, \varphi(\sqrt{2})-2\Phi(\sqrt{2})\big]^{-1} \approx 0.713, \quad
  \alpha_c(t^*) = \frac{8(t^*)^2}{\pi e^2} \approx 0.175.
\end{equation}
The solution of Eq.~\eqref{eq:pinned-spinodal} with $y\geq \sqrt{2}$ gives the disappearance of the high-overlap branch, $\alpha_c(t)$; the solution with $0<y\leq \sqrt{2}$ gives the other boundary of coexistence.
At $t=1$, the retrieval spinodal has $y \approx 2.137$ and, as expected, $\alpha_c \approx 0.138$.
For $0<t<1$, the right-hand side of Eq.~\eqref{eq:pinned-hopfield} starts at $1-t>0$ and is bounded, so there is always a positive root.
For $t<t^*$ it is unique: $\alpha(y)=[(1-t)+t g(y)]^2/y^2$ decreases from infinity to zero, since $t[y g'(y)-g(y)]<1-t$.
This proves the disappearance of the discontinuity within RS.

\bibliographystyle{elsarticle-harv}
\bibliography{bibliography}

\end{document}